\documentclass[aps,prl,twocolumn,10pt,superscriptaddress,longbibliography,floatfix]{revtex4-2}
\usepackage{amsmath,amssymb,bm,graphicx,microtype}
\usepackage[colorlinks=true,allcolors=blue]{hyperref}
\newcommand{\ii}{\mathrm{i}}
\newcommand{\Ac}{A_{\mathrm c}}
\newcommand{\pred}{\widehat A_{\mathrm c}}
\begin{document}
\title{Bypass Transition via the Non-Hermitian Skin Effect}
\author{Konghao Sun}
\affiliation{Beijing National Laboratory for Condensed Matter Physics, Institute of Physics, Chinese Academy of Sciences, Beijing 100190, China}
\author{Haiping Hu}
\email{hhu@iphy.ac.cn}
\affiliation{Beijing National Laboratory for Condensed Matter Physics, Institute of Physics, Chinese Academy of Sciences, Beijing 100190, China}
\affiliation{School of Physical Sciences, University of Chinese Academy of Sciences, Beijing 100049, China}

\begin{abstract}
Non-Hermitian skin lattices can directionally amplify propagating signals. Here we show that this amplification and nonlinear feedback together drive a bypass transition to self-sustained motion, familiar from shear flows. A finite local excitation can trigger sustained periodic or chaotic motion while the resting state remains linearly stable. We develop a return-map theory that predicts the triggering threshold, its scaling with directional gain, and the instability governing near-critical dynamics. The threshold decreases exponentially with chain length. Weak noise triggers sustained motion as well, with a characteristic ignition length that grows logarithmically as noise weakens at fixed observation time. Our work establishes a skin-effect route to bypass transitions, opening a new avenue for predicting and controlling nonlinear wave dynamics.
\end{abstract}

\maketitle

The non-Hermitian skin effect (NHSE) causes an extensive number of eigenmodes to accumulate near a boundary~\cite{HatanoNelson,Yao2018,Yokomizo2019,Okuma2020,ZhangYangFang2020,Chinghualee,Torres,Hu_NHSE,Bergholtz2021} and has been observed across diverse platforms~\cite{Weidemann2020,Helbig2020,exp2,exp3,acoustic_exp,cold_atom_exp,Gao2024Acoustic}. It can produce direction-dependent amplification or attenuation of propagating waves~\cite{Wanjura2020,Xue2021,LiWan2022,Wang2022PRB,LiQiu2026Amplification}. Potential applications include directional amplifiers~\cite{McDonald2018,Porras2019}, sensors~\cite{BudichBergholtz2020,Parto2025Sensing}, and cooling schemes~\cite{Xu2024Cooling}. Remarkably, interference between nonorthogonal eigenmodes can transiently amplify an excitation even when every mode decays~\cite{Midya2022,TrefethenEmbree2005,Makris2014,Nakai2024,Schmid2007}. Such transient growth ultimately gives way to decay without additional driving.

Nonlinearity is becoming increasingly important in skin physics~\cite{Yuce2021,ManyManda2024}, giving rise to solitons~\cite{Wang2025,Kokkinakis2026}, self-trapping~\cite{Ezawa2022}, and spatial bifurcations of stationary states~\cite{KawabataNakamura2025,Lin2026}. Nonlinear lattice dynamics can also support synchronization~\cite{Zhang2025Synchronization} and temporal chaos~\cite{Zhao2025}, and enable single-mode lasing~\cite{Zhu2022Lasing} and mode locking~\cite{Leefmans2024Laser}. In realistic systems, nonlinearity is often unavoidable, while skin amplification can magnify weak signals or fluctuations until nonlinear effects become important. A fundamental question is how such systems enter self-sustained motion, and whether this onset can be predicted quantitatively. This question is also directly relevant to the stability and performance of active devices. A predictive theory linking skin amplification to self-sustained motion is still lacking.

\begin{figure}[t]
\centering
\includegraphics[width=\linewidth]{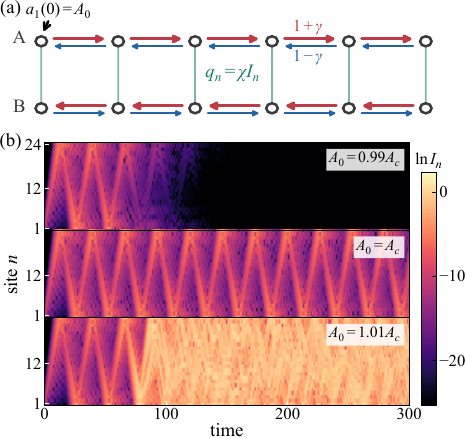}
\caption{\textbf{Bypass transition via the non-Hermitian skin effect.} (a) Two open chains A and B with opposite skin directions and uniform loss at rate $\Gamma$ on every site. Their complex field amplitudes at site $n$ are $a_n$ and $b_n$, respectively. The local conversion rate is $q_n=\chi I_n$, where $I_n=|a_n|^2+|b_n|^2$. Initially, only $a_1(0)=A_0$ is nonzero. (b) Color maps of $\ln I_n(t)$ for $A_0/A_{\mathrm c}=0.99,1,1.01$ (top to bottom), showing decay, critical circulation, and sustained chaos. Here $A_{\mathrm c}$ is the single-site triggering threshold. Parameters: $L=24$, $\gamma=0.2$, $\Gamma=0.1$, and $\chi=1$.}
\label{fig:trigger}
\end{figure}

Here we identify a bypass transition to self-sustained motion via the NHSE. A finite excitation can initiate periodic or chaotic motion while the resting state remains stable, analogous to the bypass route from a laminar state to turbulence in shear flows~\cite{Trefethen1993,GebhardtGrossmann1994,Hamilton1995,Waleffe1997,Grossmann2000}. We demonstrate the mechanism in a minimal setting of two open chains with opposite skin directions, coupled by intensity-dependent conversion [Fig.~\ref{fig:trigger}(a)]. A local excitation is amplified along one chain and converted into the other, where it is amplified again on its return. We develop a return-map theory that predicts the triggering threshold and its scaling with directional gain. The threshold decreases exponentially with chain length despite an unchanged linear decay rate. Near threshold, trajectories follow an unstable circulating orbit whose dominant instability is quantified. Weak noise can initiate sustained motion, with a characteristic triggering length that grows logarithmically as noise weakens. Our work reveals a skin-effect mechanism for sustained motion, offering a fresh perspective on nonlinear wave dynamics.

\emph{Model and finite-amplitude triggering.---}
We consider two open chains, $\mathrm A$ and $\mathrm B$, with $L$ sites each, as sketched in Fig.~\ref{fig:trigger}(a).
Their complex field amplitudes $a_n(t)$ and $b_n(t)$ obey
\begin{equation}
\begin{aligned}
\mathrm{i}\dot a_n={}&(1+\gamma)a_{n-1}+(1-\gamma)a_{n+1}+q_n b_n-\mathrm{i}\Gamma a_n,\\
\mathrm{i}\dot b_n={}&(1-\gamma)b_{n-1}+(1+\gamma)b_{n+1}+q_n a_n-\mathrm{i}\Gamma b_n.
\end{aligned}\label{eq:model}
\end{equation}
Here $n=1,\ldots,L$, with open boundaries $a_0=a_{L+1}=b_0=b_{L+1}=0$. The mean hopping is set to unity. The parameter $0<\gamma<1$ controls the hopping asymmetry, and $\Gamma>0$ denotes the uniform loss rate. The local intensity $I_n=|a_n|^2+|b_n|^2$ determines the conversion rate $q_n=\chi I_n$, with $\chi>0$. For real $q_n$, the conversion redistributes intensity between the two chains while conserving $I_n$. Unless stated otherwise, $\gamma=0.2$, $\Gamma=0.1$, and $\chi=1$.

The zero-field configuration, $a_n=b_n=0$, defines the resting state. The intensity-dependent conversion is cubic in the fields and vanishes upon linearization, leaving two independent Hatano--Nelson chains. With $\kappa=\operatorname{atanh}\gamma$, the imaginary gauge transformation $a_n=e^{\kappa(n-1)}\tilde a_n$ and $b_n=e^{-\kappa(n-1)}\tilde b_n$ maps the linearized equations to reciprocal chains with hopping $\sqrt{1-\gamma^2}$~\cite{HatanoNelson,Longhi2015}. Their eigenfrequencies are
\begin{equation}
E_m=2\sqrt{1-\gamma^2}\cos\frac{m\pi}{L+1}-\mathrm{i}\Gamma, \qquad m=1,\ldots,L.
\label{eq:spectrum}
\end{equation}
The eigenmodes accumulate at the right edge of chain $\mathrm A$ and the left edge of chain $\mathrm B$, with amplitude localization length $\kappa^{-1}$. Every eigenmode decays at the same rate $\Gamma$, independent of $L$.

\begin{figure}[t]
\centering
\includegraphics[width=\linewidth]{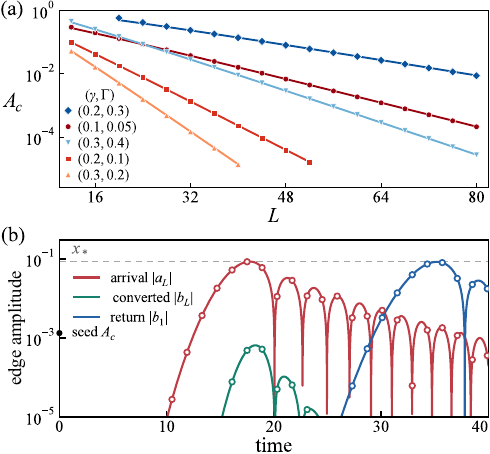}
\caption{\textbf{Return-map prediction of the triggering threshold.} (a) Single-site triggering threshold versus chain length for the parameter pairs shown. Symbols denote the numerical threshold $A_{\mathrm c}$, and lines the return-map prediction $\widehat A_{\mathrm c}$ from Eq.~\eqref{eq:threshold}. (b) Edge amplitudes $|a_L|$, $|b_L|$, and $|b_1|$ through the first passage and return at the numerical threshold for $L=32$. Solid curves show the full nonlinear dynamics and open circles the leading responses $A_{\mathrm c}a^{(1)}$ and $A_{\mathrm c}^3b^{(3)}$ from Eqs.~\eqref{eq:return_a} and \eqref{eq:return_b}. The dashed line marks the fixed-point amplitude $x_*$.}\label{fig:return}
\end{figure}

We prepare a single-site excitation by setting $a_1(0)=A_0>0$ at the leftmost site of chain $\mathrm A$, with all other amplitudes zero, and let the system evolve without further driving. For this preparation, a threshold $A_{\mathrm c}$ separates decay from sustained motion. Figure~\ref{fig:trigger}(b) illustrates the dynamics across this threshold at $L=24$. Below threshold, successive passages weaken and the lattice returns to rest. At threshold, the packet circulates repeatedly with nearly constant peak intensity over the displayed time interval. Just above threshold, irregular motion spreads throughout the lattice and persists at late times. A positive largest Lyapunov exponent confirms the chaotic character of the sustained motion~\cite{SM}. The threshold $A_{\mathrm c}$ is determined by bisection and decreases exponentially with chain length [Fig.~\ref{fig:return}(a)], while the modal decay rate remains $\Gamma$.

\emph{Return-map theory.---} To predict the triggering threshold, we calculate the field generated during the first passage and return. Since only chain $\mathrm A$ is seeded, conversion produces a field in chain $\mathrm B$ at order $A_0^3$ and feeds back into chain $\mathrm A$ at order $A_0^5$. We write
$a_n=A_0a_n^{(1)}+O(A_0^5)$ and
$b_n=A_0^3b_n^{(3)}+O(A_0^7)$.
Let $H_{\mathrm A,\mathrm B}$ be the hopping matrices and
$U_{\mathrm A,\mathrm B}(t)=e^{-\Gamma t}e^{-iH_{\mathrm A,\mathrm B}t}$
their propagators. The leading responses are
\begin{align}
a_n^{(1)}(t)
&=[U_{\mathrm A}(t)]_{n1},
\label{eq:return_a}\\
b_n^{(3)}(t)
&=-\mathrm{i}\chi\sum_{j=1}^L\int_0^t ds\,
[U_{\mathrm B}(t-s)]_{nj}
|a_j^{(1)}(s)|^2a_j^{(1)}(s).
\label{eq:return_b}
\end{align}
The first expression propagates a unit seed along chain $\mathrm A$; the second sums the cubic sources generated during the passage and propagates them through chain $\mathrm B$.

Let $G$ denote the first-arrival amplitude gain and $K$ the nonlinear return coefficient,
\begin{equation}
G=\max_t|a_L^{(1)}(t)|,
\qquad
K=\frac{\max_t|b_1^{(3)}(t)|}{G^3},
\label{eq:peaks}
\end{equation}
where the maxima are taken over the first arriving and returning packets, respectively. For a seed $A_0$, the first arrival at the right edge of chain $\mathrm A$ has peak $x_0=A_0G$, while the return to the left edge of chain $\mathrm B$ has peak $Kx_0^3$. The reflection-exchange symmetry $a_n\leftrightarrow b_{L+1-n}$ of Eq.~\eqref{eq:model} makes the two propagation legs equivalent. Approximating successive packets as having the same normalized shape up to reflection, the successive arrival peaks $x_m$ at the right edge of chain $\mathrm A$ obey
\begin{equation}
x_{m+1}\simeq K(Kx_m^3)^3=K^4x_m^9.
\label{eq:map}
\end{equation}
The map has a nonzero fixed point $x_*=K^{-1/2}$. Since $x_{m+1}/x_m=(x_m/x_*)^8$, packets below $x_*$ weaken on successive circuits, whereas those above it grow away from the fixed point. Requiring the first arrival $A_0G$ to reach $x_*$ yields
\begin{equation}
\widehat A_{\mathrm c}
=\frac{x_*}{G}
=\frac{1}{G\sqrt K}.
\label{eq:threshold}
\end{equation}

Figure~\ref{fig:return}(a) compares $\widehat A_{\mathrm c}$ from Eq.~\eqref{eq:threshold} with the thresholds $A_{\mathrm c}$ obtained from the full dynamics over several lengths and parameter sets. They agree closely across the range shown, with no fit to the threshold data. In Fig.~\ref{fig:return}(b), the leading responses $A_{\mathrm c}a^{(1)}$ and $A_{\mathrm c}^3b^{(3)}$ closely reproduce the full evolution through the first return. The converted field $|b_L|$ is roughly two orders of magnitude weaker than the arriving field $|a_L|$, but amplification along chain $\mathrm B$ regenerates a return peak $|b_1|$ of comparable height. Both the arriving and returning peaks lie close to $x_*$, as given by Eq.~\eqref{eq:threshold}.

The return map further predicts a distinctive scaling of the triggering threshold with the directional gain $G$. An arriving packet of amplitude $x$ generates a nonlinear source proportional to $\chi x^3$, while propagation along the opposite chain contributes another factor $G$. Thus $K\propto\chi G$, up to factors that vary subexponentially with $L$. It follows that $x_*\propto(\chi G)^{-1/2}$ and
\begin{equation}
\widehat A_{\mathrm c}\propto\chi^{-1/2}G^{-3/2}.
\label{eq:gainlaw}
\end{equation}
For comparison, a single-pass amplifier requires an input proportional to $G^{-1}$ to reach a fixed output amplitude. Here nonlinear regeneration also lowers the circulating amplitude needed to close the return loop, $x_*\propto G^{-1/2}$, producing the additional factor $G^{-1/2}$ in the triggering threshold. More generally, for a conversion of field order $p$ ($p=3$ here), the same argument gives
$\widehat A_{\mathrm c}\propto\chi^{-1/(p-1)}G^{-p/(p-1)}$~\cite{SM}.

The length dependence of $G$ follows from the first-passage response. This response combines the directional amplification factor $e^{\kappa(L-1)}$ with wave-packet spreading and loss. Maximizing this response gives~\cite{SM}
\begin{equation}
\ln G=r_\infty L+o(L),
\qquad
r_\infty=
\kappa-\operatorname{asinh}\!\left(
\frac{\Gamma}{2\sqrt{1-\gamma^2}}
\right).
\label{eq:response_exponent}
\end{equation}
The same exponent governs the zero-frequency end-to-end Green's function~\cite{Xue2021,LiWan2022}. Combining Eqs.~\eqref{eq:gainlaw} and \eqref{eq:response_exponent} gives
$\ln(\sqrt{\chi}\,\widehat A_{\mathrm c})=-3r_\infty L/2+o(L)$.
For $\Gamma<2\gamma$, $r_\infty>0$ and the triggering threshold falls exponentially with length, while every eigenmode still decays at rate $\Gamma$.

\begin{figure}[t]
\centering
\includegraphics[width=\linewidth]{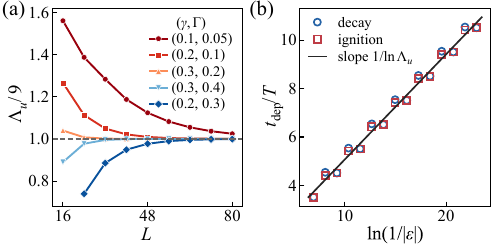}
\caption{\textbf{Critical-orbit signatures.} (a) Unstable Floquet multiplier $\Lambda_u$, normalized by the return-map prediction $\Lambda_{\rm map}=9$, versus chain length for the parameter pairs in Fig.~\ref{fig:return}(a). (b) Departure time $t_{\rm dep}/T$ near threshold at $L=24$, with $\varepsilon=A_0/A_{\mathrm c}-1$. The line uses the slope $1/\ln\Lambda_u$ fixed by the independently computed Floquet multiplier. Only the offset is fitted.}\label{fig:orbit}
\end{figure}
\emph{Critical orbit and near-threshold dynamics.---} The nonzero fixed point $x_*$ of the return map suggests a critical circulating orbit, whose instability is also predicted by the map. Writing $x_m=x_*+\delta x_m$ and expanding Eq.~\eqref{eq:map} around $x_*$ gives
\begin{equation}
\Lambda_{\rm map}
=\left.\frac{\mathrm d x_{m+1}}{\mathrm d x_m}\right|_{x_*}
=9K^4x_*^8=9.
\label{eq:multiplier}
\end{equation}
A small deviation from the fixed point is thus amplified by a factor of nine after each circuit. The coefficient $K$ cancels, leaving a factor set solely by the two successive cubic conversions. More generally, for a conversion of field order $p$, the full-circuit map has degree $p^2$ and gives $\Lambda_{\rm map}=p^2$.

We test this prediction by computing the corresponding periodic orbit independently from Eq.~\eqref{eq:model} using Newton shooting~\cite{SM}. Its period $T$ is the time between successive arrivals at the same edge, and its arriving peak agrees closely with $x_*$. Its stability is characterized by the Floquet multipliers~\cite{GuckenheimerHolmes1983}; apart from the neutral directions, there is a single unstable multiplier $\Lambda_u>1$. Figure~\ref{fig:orbit}(a) shows that $\Lambda_u$ approaches the return-map prediction $9$ with increasing length across the parameter sets considered. The threshold trajectory approaches this orbit, while nearby trajectories follow it temporarily before departing toward decay or sustained motion. The critical orbit thus lies on the boundary between the two outcomes, analogous to unstable solutions on the laminar--turbulent boundary in shear flows~\cite{Skufca2006,Schneider2007,Kawahara2012}.

For a near-threshold seed $A_0=A_{\mathrm c}(1+\varepsilon)$, the unstable displacement is proportional to $\varepsilon$ and grows after $m$ circuits as $|\varepsilon|\Lambda_u^m$. Defining $t_{\rm dep}$ as the departure time from the critical orbit gives
\begin{equation}
\frac{t_{\rm dep}}{T}
\simeq c_0+\frac{\ln(1/|\varepsilon|)}{\ln\Lambda_u},
\label{eq:departure}
\end{equation}
where $c_0$ is an offset.
Figure~\ref{fig:orbit}(b) confirms this logarithmic dependence on both sides of threshold. The independently computed $\Lambda_u$ fixes the slope, while only $c_0$ is fitted. The logarithmic delay reflects repeated circuits near the unstable orbit rather than a slowing of the resting-state decay.

\emph{Power balance and the feedback loop.---} The local conversion redistributes intensity between the chains but supplies no net power. The total power $P=\sum_n I_n$ obeys
\begin{equation}
\dot P=4\gamma\sum_{n=1}^{L-1}\operatorname{Im}\!\left(b_n^*b_{n+1}-a_n^*a_{n+1}\right)-2\Gamma P.
\label{eq:power}
\end{equation}
In sustained motion, the power supplied by nonreciprocal hopping balances the loss on average. In the chaotic regime, the mean power grows approximately linearly with $L$ and depends only weakly on the above-threshold seed amplitude~\cite{SM}. The seed triggers the motion but does not power it.

For sufficiently strong loss, nonreciprocal hopping can no longer sustain the field. Equation~\eqref{eq:power} gives a global decay bound: every finite open lattice decays for $\Gamma\ge2\gamma$, regardless of the initial amplitude~\cite{SM}. This boundary also has a simple bulk interpretation. The periodic dispersions $E_{\mathrm A,\mathrm B}(k)=2\cos k-\mathrm{i}\Gamma\mp2\mathrm{i}\gamma\sin k$ have maximum growth rate $2\gamma-\Gamma$, so $\Gamma<2\gamma$, equivalently $r_\infty>0$, is the regime of directional amplification~\cite{Wanjura2020,Xue2021,Midya2022}, even though all open-boundary eigenmodes decay. Directional amplification alone, however, cannot sustain motion. If both chains amplify toward the same end, the resting-state spectrum and optimal transient gain remain unchanged, yet the weighted power $P_W=\sum_n e^{-2\kappa(n-1)}I_n$ obeys $\dot P_W=-2\Gamma P_W$, forcing every field to decay. With opposite skin directions, conversion closes the return loop by routing the field into another amplifying channel.

\begin{figure}[t]
\centering
\includegraphics[width=\linewidth]{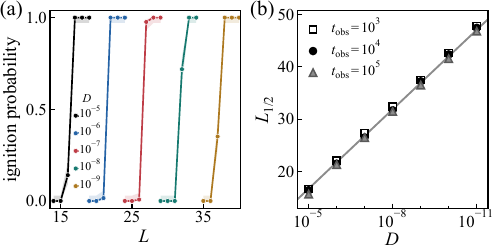}
\caption{\textbf{Noise-controlled ignition length.} (a) Ignition probability versus chain length $L$ for independent complex Gaussian white noise applied to each field component, with observation time $t_{\rm obs}=10^4$. Ignition is recorded when $P>1$. Each point represents $128$ realizations, with shading indicating $95\%$ confidence intervals. (b) Half-ignition length $L_{1/2}$ versus noise strength $D$ for the observation times shown. The line follows Eq.~\eqref{eq:LD}, with only the offset fitted to the $t_{\rm obs}=10^4$ data.}\label{fig:noise}
\end{figure}
\emph{Noise-induced ignition.---} Noise can also supply the finite disturbance required to trigger sustained motion. Starting from the resting state, we add independent complex Gaussian white noise $\eta_j(t)$ to each of the $2L$ field components, with
$\langle\eta_j(t)\eta_k^*(t')\rangle=2D\delta_{jk}\delta(t-t')$,
where $D$ is the noise strength. We record whether ignition occurs within an observation time $t_{\rm obs}$. Figure~\ref{fig:noise}(a) shows a sharp crossover with chain length: at fixed $D$, the ignition probability rises from nearly zero to nearly one over only a few sites, while weaker noise shifts the crossover to longer chains. Once triggered, the motion persists after the noise is removed~\cite{SM}.

Directional amplification controls this strong length dependence. Fluctuations generated near the left edge of chain $\mathrm A$ and the right edge of chain $\mathrm B$ experience the longest amplified passage and dominate the opposite-edge variance~\cite{SM}. To leading exponential order, the critical circulating intensity scales as $x_*^2=(\widehat A_{\mathrm c}G)^2$, while the endpoint noise variance scales as $DG^2$. Their ratio scales as $\widehat A_{\mathrm c}^2/D$.

Let $L_{1/2}$ denote the chain length at which half the realizations ignite within $t_{\rm obs}$. At fixed $t_{\rm obs}$, we estimate the ignition probability primarily from this ratio. The half-ignition condition then corresponds approximately to a fixed value of $\widehat A_{\mathrm c}^2/D$. Together with $\widehat A_{\mathrm c}^2\propto\chi^{-1}e^{-3r_\infty L}$, this gives
\begin{equation}
L_{1/2}(D;t_{\rm obs})\simeq\frac{1}{3r_\infty}\ln\frac{1}{\chi D}+C(t_{\rm obs}).
\label{eq:LD}
\end{equation}
Figure~\ref{fig:noise}(b) confirms the predicted logarithmic dependence, with only the offset $C(t_{\rm obs})$ fitted. Changing $t_{\rm obs}$ shifts the crossover much less than changing the noise strength. The same directional gain that lowers the deterministic threshold also amplifies noise, so reducing the noise increases the characteristic ignition length only logarithmically.

\emph{Discussion.---} We have shown that NHSE-induced amplification and nonlinear feedback can produce a bypass transition to self-sustained motion. The triggering threshold falls exponentially with system size, while a return-map theory captures both the threshold and the near-critical dynamics. The transition concerns the route into sustained motion rather than the final attractor, which may be chaotic or periodic~\cite{SM}. Notably, the triggering process is tractable in the skin setting: for the specified local excitation, the spatially organized amplification and nonlinear return can be calculated explicitly, without nonlinear optimization or edge tracking~\cite{PringleKerswell2010,Kerswell2018}. More broadly, our results reveal skin amplification not merely as a transient response, but as a dynamical resource that nonlinear feedback can repeatedly harness.

The mechanism involves transient directional amplification, nonlinear regeneration, and an energy source that balances dissipation. These ingredients are structural rather than specific to the minimal two-chain model: the same mechanism survives, e.g., with unequal propagation channels, dimerized lattices, additional Kerr nonlinearities, and different nonlinear orders~\cite{SM}. This points to a broader class of nonreciprocal and nonnormal systems in which transient amplification can be converted into autonomous dynamics. The predicted thresholds and scaling laws could be tested in active electrical networks, where nonreciprocal couplings~\cite{Helbig2020,LiuCircuit2021} and nonlinear circuit elements~\cite{Hadad2018} are available. For active non-Hermitian devices, these results provide guidance for stability and nonlinear performance, and may also inform the design of switches and locally triggered oscillators.


\begin{acknowledgments}
This work is supported by the National Key Research and Development Program of China (Grants No. 2023YFA1406704 and No. 2022YFA1405800) and National Natural Science Foundation of China (Grant No. 12474496 and No. 12547107).
\end{acknowledgments}

\label{maintext:end}

\onecolumngrid
\clearpage
\setcounter{section}{0}
\setcounter{subsection}{0}
\setcounter{equation}{0}
\setcounter{figure}{0}
\setcounter{table}{0}
\setcounter{secnumdepth}{2}
\renewcommand{\ii}{\mathrm i}
\newcommand{\Id}{\mathbb I}
\renewcommand{\thesection}{S\arabic{section}}
\renewcommand{\theequation}{S\arabic{equation}}
\renewcommand{\thefigure}{S\arabic{figure}}
\renewcommand{\thetable}{S\arabic{table}}
\renewcommand{\theHsection}{supp.\arabic{section}}
\renewcommand{\theHequation}{supp.\arabic{equation}}
\renewcommand{\theHfigure}{supp.\arabic{figure}}
\renewcommand{\theHtable}{supp.\arabic{table}}
\phantomsection
\label{sm:start}

\section*{Supplemental Material for ``Bypass Transition via the Non-Hermitian Skin Effect''}
\raggedbottom

This Supplement specifies the numerical methods, derives the return prediction and its length dependence, and connects it to the critical circulating orbit. It then examines the power source and decay controls, noise-induced ignition, periodic sustained states, dimerized chains, and residual linear coupling.

\begingroup
\makeatletter
\let\sm@contentsline\contentsline
\def\contentsline#1#2#3#4{%
  \def\sm@kind{#1}\def\sm@section{section}%
  \ifx\sm@kind\sm@section
    \sm@numberedentry#2\@nil{#3}{#4}%
  \fi
}
\def\sm@numberedentry\numberline#1#2\@nil#3#4{%
  \if\relax\detokenize{#1}\relax\else
    \sm@contentsline{section}{\numberline{#1}#2}{#3}{#4}%
  \fi
}
\def\l@f@section{\addpenalty{\@secpenalty}\addvspace{0.25em}}
\tableofcontents
\makeatother
\endgroup

\section{Preparation and numerical methods}\label{sm:sec:methods}

We prepare $a_1(0)=A_0\ge0$ with all other fields zero and let the lattice evolve without further driving. Unless stated otherwise, $\gamma=0.2$, $\Gamma=0.1$, and $\chi=1$. Along this family of initial states, $A_{\mathrm c}$ denotes the first threshold between decay and sustained motion, which may be chaotic or regular. We use $\pred$ for the threshold predicted by the return condition.

The real single-site seed belongs to the fixed subspace of the antilinear symmetry
\begin{equation}
\mathcal T(a_n,b_n)=\bigl((-1)^{n+1}a_n^*,\,(-1)^nb_n^*\bigr).
\label{sm:eq:preparation_symmetry}
\end{equation}
Within this subspace, $a_n$ is real on odd sites and imaginary on even sites, with the opposite pattern for $b_n$. Hopping and conversion connect fields of opposite real--imaginary type, so the factor $-\ii$ makes each time derivative match the type of its field. Uniform loss preserves this pattern as well, so a trajectory starting in the subspace remains there. We use this same single-site preparation for thresholds, power statistics, and periodic states, while testing stability against perturbations both within and outside the subspace.

\paragraph{Deterministic calculations.}
Deterministic threshold scans use fourth-order Runge--Kutta integration with $\Delta t=0.02$ and bisection between decay and sustained motion. Runs initially extend to $t=10^4$. We sample the total power once per unit time and average it in blocks of duration $100$ over the final $10^3$ units. A trajectory is classified as decaying if every block mean is below $10^{-12}P(0)$ and as sustained if every block mean exceeds $0.05$. Otherwise, the integration time is doubled before assigning an outcome. Bisection reduces the relative threshold bracket to below $10^{-5}$. For the near-threshold departure times, we refine it to below $10^{-13}$, well below the smallest relative seed displacement used. Halving the step and doubling the duration preserve the outcomes just below and above threshold.

\paragraph{Response and stability calculations.}
We integrate the response equations with the adaptive eighth-order Runge--Kutta method DOP853, using relative and absolute tolerances of $2\times10^{-10}$ and $10^{-11}$ and sampling endpoint fields every $0.02$ time units. Floquet multipliers and Lyapunov spectra~\cite{Benettin1980} are obtained from the full-field variational equations, as described in Secs.~\ref{sm:sec:orbit} and \ref{sm:sec:chaos}.

\section{From propagation to the triggering threshold}\label{sm:sec:return}

\paragraph{The initial field and its linear propagation.}

Writing the site amplitudes as
$\bm a=(a_1,\ldots,a_L)^{\mathsf T}$ and
$\bm b=(b_1,\ldots,b_L)^{\mathsf T}$, the equations of the Letter become
\begin{align}
\dot{\bm a}&=M_A\bm a-\ii\chi\bm I\cdot\bm b,
&\dot{\bm b}&=M_B\bm b-\ii\chi\bm I\cdot\bm a,\label{sm:eq:model}\\
I_n&=|a_n|^2+|b_n|^2,
&M_s&=-\ii H_s-\Gamma\Id_L .\notag
\end{align}
Here $\bm I=(I_1,\ldots,I_L)^{\mathsf T}$ contains the local intensities.
The dot denotes multiplication at each site,
$(\bm u\cdot\bm v)_n=u_nv_n$. Inner products are written as
$\bm u^\dagger\bm v$. The matrix $M_s$ describes hopping and loss in
chain $s=A,B$, and $\Id_L$ is the $L\times L$ identity.
The hopping matrix $H_A$ has entries $(H_A)_{n+1,n}=1+\gamma$ and
$(H_A)_{n,n+1}=1-\gamma$, with all other entries zero.
The opposite hopping direction in chain $B$ gives $H_B=H_A^{\mathsf T}$.
We label site $n$ of chain $A$ ($B$) by $A_n$ ($B_n$).
The single-site preparation is $\bm a(0)=A_0\bm e_1$ and $\bm b(0)=0$,
where $\bm e_1=(1,0,\ldots,0)^{\mathsf T}$. Its total power is
$P(0)=A_0^2$, with $P=\|\bm a\|^2+\|\bm b\|^2$.

The conversion vanishes at linear order about the resting state,
leaving two independently propagating fields. Their propagators are
$U_s(t)=e^{M_st}$, so that
$\bm a(t)=U_A(t)\bm a(0)$ and $\bm b(t)=U_B(t)\bm b(0)$.
The matrix element $[U_s(t)]_{nj}$ is the field at site $n$ and time $t$
produced by a unit initial field at site $j$ in chain $s$.
For the single-site preparation, this gives
$a_n(t)=A_0[U_A(t)]_{n1}$, while chain $B$ remains empty.

To evaluate these propagators, we use the imaginary gauge transformation
$\bm a=S\widetilde{\bm a}$ and $\bm b=S^{-1}\widetilde{\bm b}$,
where $S_{nn}=e^{\kappa(n-1)}$ and $\kappa=\operatorname{atanh}\gamma$.
Since $(1-\gamma)e^\kappa=(1+\gamma)e^{-\kappa}=h$, with
$h=\sqrt{1-\gamma^2}$, both transformed chains have symmetric hopping,
$S^{-1}H_AS=SH_BS^{-1}=hH_0$. Here $H_0$ is the open chain with unit
nearest-neighbor hopping in both directions. Its standing waves are
proportional to $\sin(k_mn)$, giving the eigenvalues of $M_A$ and $M_B$
as~\cite{HatanoNelson,Yokomizo2019}
\begin{equation}
\lambda_m=-\Gamma-2\ii h\cos k_m,\qquad
k_m=\frac{m\pi}{L+1},\qquad m=1,\ldots,L.
\label{sm:eq:linear_spectrum}
\end{equation}
These growth rates satisfy $\lambda_m=-\ii E_m$, where $E_m$ are the
mode frequencies in the Letter.
Expanding in these standing waves and restoring the gauge factors gives
\begin{equation}
[U_A(t)]_{nj}=e^{\kappa(n-j)-\Gamma t}\frac{2}{L+1}
\sum_{m=1}^L\sin(k_mn)\sin(k_mj)e^{-2\ii ht\cos k_m}.
\label{sm:eq:kernel}
\end{equation}
Replacing $\kappa$ by $-\kappa$ gives $U_B(t)$.
The factor $e^{\kappa(n-j)}$ favors propagation toward increasing $n$
in chain $A$, while its inverse favors the opposite direction in chain $B$.
Both propagators also contain the uniform decay factor $e^{-\Gamma t}$.

\paragraph{The amplified field becomes a source in the other chain.}

Conversion couples these propagating fields by supplying a local source
to each chain. Applying the variation-of-constants formula to
Eq.~\eqref{sm:eq:model}, with $\bm a(0)=A_0\bm e_1$ and $\bm b(0)=0$,
gives the exact relations
\begin{align}
\bm a(t)&=A_0U_A(t)\bm e_1-\ii\chi\int_0^t U_A(t-s)[\bm I\cdot\bm b](s)\,ds,\\
\bm b(t)&=-\ii\chi\int_0^t U_B(t-s)[\bm I\cdot\bm a](s)\,ds.\label{sm:eq:duhamel}
\end{align}
The integrals propagate each source from its generation time $s$ to
the observation time $t$. Since the initial field in chain $A$ is of
order $A_0$ and its intensity is of order $A_0^2$, the source
$-\ii\chi I_na_n$ first generates a field of order $A_0^3$ in chain $B$.
Feeding this field back into chain $A$ gives a correction of order
$A_0^2A_0^3=A_0^5$.
We therefore expand
\begin{equation}
\bm a=A_0\bm a^{(1)}+A_0^5\bm a^{(5)}+\cdots,\qquad
\bm b=A_0^3\bm b^{(3)}+O(A_0^7).
\end{equation}
The superscripts label orders in the seed amplitude, not powers of the
fields. Substituting this expansion into Eq.~\eqref{sm:eq:model} gives
the leading response equations
\begin{align}
\dot{\bm a}^{(1)}&=M_A\bm a^{(1)},&\bm a^{(1)}(0)&=\bm e_1,\label{sm:eq:linear_aux}\\
\dot{\bm b}^{(3)}&=M_B\bm b^{(3)}-\ii\chi|\bm a^{(1)}|^2\cdot\bm a^{(1)},&\bm b^{(3)}(0)&=0.\label{sm:eq:cubic_aux}
\end{align}
The first equation propagates the unit seed,
$\bm a^{(1)}(t)=U_A(t)\bm e_1$. Substituting it into the source for
chain $B$ gives
\begin{equation}
b_n^{(3)}(t)=-\ii\chi\sum_{j=1}^L\int_0^t[U_B(t-s)]_{nj}
|a_j^{(1)}(s)|^2a_j^{(1)}(s)\,ds.\label{sm:eq:cubic_return}
\end{equation}
The return is the coherent sum of the fields generated along chain $A$
and propagated back through chain $B$ by $U_B(t-s)$.
For a seed of amplitude $A_0$, the leading return field is
$A_0^3\bm b^{(3)}$. The conversion strength $\chi$ is already included
in $\bm b^{(3)}$.

\paragraph{Closing the return.}

Let $G$ be the first arrival peak of the unit-seed response at $A_L$,
and $Y$ the first return peak of the cubic response at $B_1$,
\begin{equation}
G=\max_{t\in W_o}|a_L^{(1)}(t)|,\qquad
Y=\max_{t\in W_r}|b_1^{(3)}(t)|.\label{sm:eq:peaks}
\end{equation}
The windows $W_o$ and $W_r$ select the first arriving and returning
packets, respectively. The endpoint coefficient $Y$ includes sources
throughout the chain, as specified by Eq.~\eqref{sm:eq:cubic_return}.

The seed produces an arrival peak $x=A_0G$ and a return peak $y=A_0^3Y$.
With $K=Y/G^3$, the return is related to the arrival by $y=Kx^3$.
We extend this relation to successive passages by approximating the
arriving packets as having the same normalized shape.
Reflection and exchange of the two identical chains then give the same
coefficient $K$ for both directions. Two successive conversions give
\begin{equation}
y=Kx^3,\qquad x'=Ky^3=K^4x^9.\label{sm:eq:peak_map}
\end{equation}
Here $x'$ is the next arrival at $A_L$, after the field has returned
through chain $B$ and traveled along chain $A$ again.
A packet that repeats without growing or shrinking satisfies $x'=x$.
Its nonzero peak $x_*$ and the seed needed to reach it are therefore
\begin{equation}
x_*=K^{-1/2},\qquad \pred=\frac{x_*}{G}=\frac{1}{G\sqrt K}=\sqrt{\frac{G}{Y}}.\label{sm:eq:prediction}
\end{equation}
Since $x'/x=(x/x_*)^8$, peaks below $x_*$ decrease on successive circuits
and those above it grow. Equivalently, $\pred$ is the seed for which
$A_0^3Y=A_0G$, so the returning peak equals the arriving peak.
The response coefficients $G$ and $Y$ thus determine the predicted
threshold without a nonlinear threshold search. Its circulating
amplitude and instability are compared with the full-lattice periodic
orbit in Sec.~\ref{sm:sec:orbit}.

In the numerical response calculation, we take
$W_o=[0.4,1.7]\tau_L$ and $W_r=[1.1,3.2]\tau_L$.
The reference transit time $\tau_L=(L-1)/(2h)$ is the end-to-end
distance divided by the maximum group velocity $2h$ of the reciprocal
chain obtained by the imaginary gauge transformation.

\paragraph{Gain scaling and comparison with simulations.}

The arriving field has amplitude $x=A_0G$, so it supplies a cubic source
of size $\chi x^3=\chi A_0^3G^3$. Propagation back along the mirror-related
chain contributes one further gain factor $G$, giving
$y\sim\chi A_0^3G^4$ and hence $Y\sim\chi G^4$.
The linear dependence on $\chi$ is exact in the cubic response, while $G^4$
estimates the leading amplification.
The long-chain gain, derived below by maximizing the linear response, is
\begin{equation}
r_\infty=\kappa-\operatorname{asinh}\frac{\Gamma}{2h},\qquad
\ln G=r_\infty L+o(L).\label{sm:eq:rinf}
\end{equation}
When conversion is dominated by the ends and source contributions do
not cancel at leading exponential order, the four gain factors give
\begin{equation}
G=e^{r_\infty L+o(L)},\qquad
Y=\chi e^{4r_\infty L+o(L)},\qquad
\pred=\chi^{-1/2}e^{-3r_\infty L/2+o(L)}.\label{sm:eq:asymptote}
\end{equation}
Here $o(L)$ denotes corrections that grow more slowly than $L$,
including the logarithms of algebraic spreading factors. They account
for the difference between finite-length slopes and the asymptotic
slope $3r_\infty/2$. Unlike the length asymptote, the dependence on
$\chi$ is exact at every $L$. Multiplying all fields by $\sqrt\chi$
reduces the full equations to those with $\chi=1$, so thresholds scale
as $\chi^{-1/2}$ and corresponding powers as $\chi^{-1}$.

The corrections to the gain law can be collected in a single factor,
\begin{equation}
C_{\rm pkt}=\left(\frac{\chi G^4}{Y}\right)^{1/2},\qquad
\pred=\frac{C_{\rm pkt}}{\sqrt\chi\,G^{3/2}}.\label{sm:eq:prefactor}
\end{equation}
This rewrites Eq.~\eqref{sm:eq:prediction} exactly, with
$C_{\rm pkt}$ calculated from the responses rather than fitted to the
threshold. It accounts for packet shape and conversion over space and time.

The same argument applies to conversion of field order $p>1$,
$q_n=\chi I_n^{(p-1)/2}$. An arriving field of amplitude $x$
generates a source of order $\chi x^p$. Return propagation gives
$y=K_px^p$, with $K_p\propto\chi G$ to leading exponential order.
For identical chains, the circulating peak satisfies $K_px_*^p=x_*$,
so
\begin{equation}
x_*=K_p^{-1/(p-1)},\qquad
\pred=\frac{x_*}{G}
\propto\chi^{-1/(p-1)}G^{-p/(p-1)}.
\label{sm:eq:p_gainlaw}
\end{equation}
Within the same packet-shape approximation, the conversion order
therefore sets the power of the gain, while propagation sets its
exponential dependence on length.

\begin{figure}[!htbp]
\includegraphics[width=.8\linewidth]{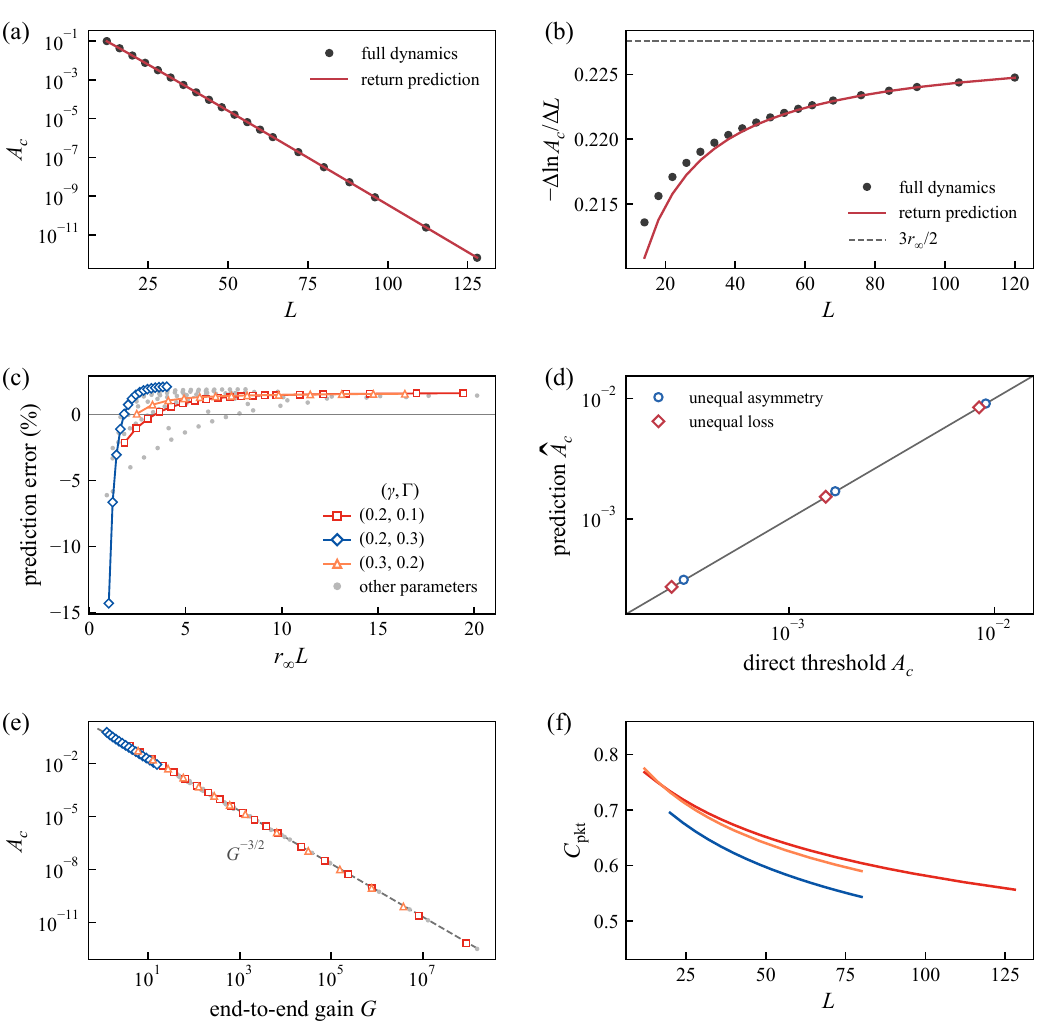}
\caption{Single-site threshold prediction and gain scaling. (a) Nonlinear thresholds and endpoint predictions at the reference parameters. (b) Local length slopes, with asymptote $3r_\infty/2$. Symbols and solid curves in (a,b) denote nonlinear calculations and response predictions. (c) Signed relative prediction error, $100(\pred/\Ac-1)\%$. (d) Unequal-chain tests of Eq.~\eqref{sm:eq:unequal}, at $L=24,32,40$. The diagonal denotes equality. (e) Nonlinear thresholds versus gain $G$. The dashed line indicates the $-3/2$ slope. (f) Response factor $C_{\rm pkt}(L)$ from Eq.~\eqref{sm:eq:prefactor}. The same colors identify three parameter pairs in (c,e,f). Gray points in (c,e) show the other eight pairs listed in Table~\ref{sm:tab:parameters}.}
\label{sm:fig:validation}
\end{figure}

At the reference parameters, the thresholds span eleven orders of
magnitude up to $L=128$, with local length slopes approaching
$3r_\infty/2$ from below [Fig.~\ref{sm:fig:validation}(a,b)].
Across the eleven parameter pairs in Fig.~\ref{sm:fig:validation}(c),
the largest prediction errors occur at small $r_\infty L$.
This includes short chains and parameters near the amplification
boundary $r_\infty=0$. Stronger gain lowers the required seed and
improves the first-passage approximation.

Figure~\ref{sm:fig:validation}(e) compares the thresholds with the leading
$-3/2$ gain dependence across all eleven parameter pairs.
The response factor $C_{\rm pkt}$ varies more slowly than the exponential
gain [Fig.~\ref{sm:fig:validation}(f)], as required for the leading
exponential law. Table~\ref{sm:tab:parameters} lists its finite-length
ranges together with the prediction errors.

\begin{table}[t]
\centering
\begin{minipage}{.8\linewidth}
\caption{Parameter coverage of the single-endpoint prediction. $C_{\rm pkt}$ is calculated from the response coefficients using Eq.~\eqref{sm:eq:prefactor}. The error column is the largest absolute relative error within the listed length range.}
\label{sm:tab:parameters}
\begin{ruledtabular}
\begin{tabular}{ccccc}
$\gamma$ & $\Gamma$ & $L$ & maximum error (\%) & range of $C_{\rm pkt}$ \\
0.1 & 0.05 & 12--80 & 6.1 & 0.57--0.70 \\
0.2 & 0.05 & 12--64 & 4.0 & 0.64--0.77 \\
0.2 & 0.1 & 12--128 & 2.1 & 0.56--0.77 \\
0.2 & 0.15 & 12--64 & 1.6 & 0.61--0.76 \\
0.2 & 0.2 & 12--64 & 2.5 & 0.60--0.76 \\
0.2 & 0.3 & 20--80 & 14.3 & 0.54--0.70 \\
0.3 & 0.1 & 16--64 & 1.3 & 0.64--0.76 \\
0.3 & 0.2 & 12--80 & 1.6 & 0.59--0.77 \\
0.3 & 0.3 & 16--64 & 1.7 & 0.59--0.74 \\
0.3 & 0.4 & 12--80 & 5.8 & 0.54--0.76 \\
0.4 & 0.2 & 16--64 & 1.4 & 0.61--0.75 \\
\end{tabular}
\end{ruledtabular}
\end{minipage}
\end{table}

\newpage
\paragraph{Unequal chains.}

If the chains have different hopping asymmetries $\gamma_A,\gamma_B>0$
or losses $\Gamma_A,\Gamma_B$, their propagation directions are still
opposite but the two return coefficients need not be equal.
The response equations determine each coefficient separately.
A unit seed at $A_1$ gives an arrival $G_A$ at $A_L$ and a return $Y_B$
at $B_1$, defining $K_B=Y_B/G_A^3$.
For the reverse passage, a unit seed at $B_L$ gives an arrival $G_B$
at $B_1$ and a return $Y_A$ at $A_L$, defining $K_A=Y_A/G_B^3$.
These auxiliary unit-seed responses are used to predict the threshold
for the original preparation, $a_1(0)=A_0$ with all other fields zero.
Applying the two return relations in sequence gives
\begin{equation}
y=K_Bx^3,\qquad x'=K_Ay^3,\qquad
x_*=(K_AK_B^3)^{-1/8},\quad y_*=(K_BK_A^3)^{-1/8}.\label{sm:eq:unequal}
\end{equation}
The circulating peaks $x_*$ and $y_*$ are obtained by imposing $x'=x$ and
$y=K_Bx^3$. The predicted threshold is again the seed whose first arrival
reaches $x_*$, now $\pred=x_*/G_A$.
Starting from the reference parameters, we either reduce $\gamma_B$
to $0.18$ or increase $\Gamma_B$ to $0.12$, and test $L=24,32,40$.
The resting-state decay margin remains $\min(\Gamma_A,\Gamma_B)=0.1$
independently of length. The predictions differ from the numerical
thresholds by at most $1.21\%$ [Fig.~\ref{sm:fig:validation}(d)], with
threshold brackets checked as in Sec.~\ref{sm:sec:methods}.
For identical chains, $K_A=K_B=K$ and $G_A=G$, recovering the
symmetric-chain result $\pred=K^{-1/2}/G=\sqrt{G/Y}$.

\paragraph{The long-chain gain exponent.}\label{sm:sec:exponent}

Equation~\eqref{sm:eq:kernel} gives the gain $G$ at finite length.
Its large-length exponent is most easily obtained from the corresponding
infinite-chain propagator $U(d,t)$, where $d=n-j>0$ is the distance from
a unit seed at site $j$ to the observation site $n$.
In the gauge-transformed chain, a single-site seed has equal Fourier
weight at all wave numbers, whose dispersion is $2h\cos k$.
Adding the propagation factors $e^{\ii kd-2\ii ht\cos k}$ and restoring
the gauge factor $e^{\kappa d}$ and loss $e^{-\Gamma t}$ gives
\begin{equation}
U(d,t)=\frac{e^{\kappa d-\Gamma t}}{2\pi}\int_{-\pi}^{\pi}e^{\ii kd-2\ii ht\cos k}\,dk.
\end{equation}
At fixed distance $d$, the directional factor $e^{\kappa d}$ is constant,
so the peak occurs where reciprocal propagation and loss combine to
give the largest response.
At and after the reciprocal chain's ballistic arrival,
$t\ge d/(2h)$, the unit modulus of the integrand on the real axis gives
\[
|U(d,t)|\le e^{\kappa d-\Gamma t}
\le \exp\!\left[\left(\kappa-\frac{\Gamma}{2h}\right)d\right],
\qquad t\ge\frac{d}{2h}.
\]
The earlier tail suffers less loss and can exceed this late-time bound.
For $0<t<d/(2h)$ and large $d$, contributions from different wave numbers
largely cancel as their phases vary rapidly. The remaining response
can be evaluated by steepest descent~\cite{Longhi2019,XueEdge2026}.
Writing the exponent as $\Phi(k)=\ii kd-2\ii ht\cos k$, its stationary
points satisfy
\[
\Phi'(k_*)=\ii d+2\ii ht\sin k_*=0,
\qquad \sin k_*=-\frac{d}{2ht}.
\]
Before the ballistic arrival, this equation has no real solution.
The early-time tail is governed by the complex saddle
\[
k_*=-\frac{\pi}{2}+\ii\alpha,
\qquad \cosh\alpha=\frac{d}{2ht},\qquad \alpha>0.
\]
The integration path can be shifted upward by $\ii\alpha$ to pass through
this saddle. No singularities are crossed, and the vertical sides of the
connecting contour cancel because the integrand is $2\pi$-periodic for
integer $d$. Cauchy's theorem therefore leaves the integral unchanged.
On the shifted path $k=k_*+u$, with real $u$, the integrand's modulus
falls away from $u=0$ by the factor
$\exp[-2ht\sinh\alpha(1-\cos u)]$.
Near the saddle,
\[
\Phi(k_*+u)=\Phi(k_*)-ht\sinh\alpha\,u^2+O(d\,u^3).
\]
For large $d$ at fixed $0<t/d<1/(2h)$, the contributing region has
width of order $d^{-1/2}$. Across this narrow region, the phase changes
only by order $d\,u^3=O(d^{-1/2})$, so the contributions add nearly in phase.
The rest of the path is exponentially suppressed, so this Gaussian
neighborhood dominates the integral. Its width supplies an algebraic
prefactor, while the leading exponential dependence is set by
$\operatorname{Re}\Phi(k_*)=-d\alpha+2ht\sinh\alpha$, so
\begin{equation}
\ln|U(d,t)|=\kappa d-\Gamma t-d\alpha+2ht\sinh\alpha+o(d).
\label{sm:eq:ray}
\end{equation}
The prefactor contributes only to the $o(d)$ term.

The peak follows by maximizing Eq.~\eqref{sm:eq:ray} over time at fixed $d$.
Although $\alpha$ depends on $t$, the terms containing $d\alpha/dt$
cancel because $2ht\cosh\alpha=d$, leaving
\[
\frac{d}{dt}\left[-\Gamma t-d\alpha+2ht\sinh\alpha\right]
=-\Gamma+2h\sinh\alpha
=-\Gamma+\sqrt{\frac{d^2}{t^2}-4h^2}.
\]
The derivative decreases from $+\infty$ to $-\Gamma$, giving a
unique maximum at
\[
\alpha_*=\operatorname{asinh}\frac{\Gamma}{2h},
\qquad t_{\rm peak}\simeq\frac{d}{\sqrt{\Gamma^2+4h^2}}.
\]
At the peak, $2ht\sinh\alpha_*=\Gamma t$, so the exponent reduces
to $d(\kappa-\alpha_*)$.
For $\Gamma>0$, $\alpha_*<\Gamma/(2h)$, making this exponent larger
than the late-time upper bound. The early peak therefore gives the
global maximum at leading exponential order,
$\ln\max_{t>0}|U(d,t)|=d(\kappa-\alpha_*)+o(d)$.
This recovers $r_\infty$ in Eq.~\eqref{sm:eq:rinf}. At finite distance,
prefactors can shift the peak time without changing its asymptotic exponent.

To impose open boundaries, we add image sources to the gauge-transformed
propagator so that the field vanishes at the two fictitious sites
$0$ and $L+1$.
The leading terms near the first arrival have propagation distances
$L-1$, $L+1$, and $L+3$, which differ only by fixed offsets and therefore
share the same large-$L$ exponent.
Their sum retains this exponent at fixed $\Gamma>0$, while contributions
from additional traversals are exponentially smaller at these times.
The boundaries thus change the prefactor, not $r_\infty$.
For the finite-length value of $G$, we use Eq.~\eqref{sm:eq:kernel}, which
retains these boundary effects and packet spreading.

The exponent $r_\infty$ also appears in the fixed-frequency response.
For a real frequency $\omega$, let $\beta$ be the complex amplitude
ratio between neighboring sites. Substituting
$a_n=\beta^ne^{-\ii\omega t}$ into the linear bulk equation gives
\begin{equation}
(1-\gamma)\beta^2-(\omega+\ii\Gamma)\beta+(1+\gamma)=0.
\end{equation}
At the temporal peak found above, the saddle satisfies
$2h\cos k_*-\ii\Gamma=0$.
Restoring the gauge factor gives
$\beta=e^{\kappa+\ii k_*}=-\ii e^{\kappa-\alpha_*}$,
the smaller-magnitude root at $\omega=0$.
Thus maximizing the pulse response selects the same spatial exponent
$\ln|\beta_<|=r_\infty$ as the zero-frequency
response~\cite{Xue2021,LiWan2022}.

\section{The critical circulating orbit}\label{sm:sec:orbit}

\paragraph{A full-field solution and its instability.}

The nonzero fixed point of the return map describes a packet that
reproduces its peak after each circuit. To find the corresponding orbit
of the full lattice, let $\psi=(\bm a,\bm b)$ denote the field at one
instant and $\Phi_T(\psi)$ its value after evolving Eq.~\eqref{sm:eq:model}
for a time $T$. An orbit that repeats up to a common phase $\theta$ satisfies
\begin{equation}
\Phi_T(\psi)-e^{\ii\theta}\psi=0.\label{sm:eq:shooting}
\end{equation}
A near-threshold trajectory provides an initial guess for this orbit.
We take a reference field $\psi_{\rm ref}$ from one arrival and estimate
$T$ from the interval between successive, nearly identical arrivals.
Evaluating Eq.~\eqref{sm:eq:model} at $\psi_{\rm ref}$ gives its time
derivative $F_{\rm ref}$. Small changes of time origin and common phase
move this field along $F_{\rm ref}$ and $\ii\psi_{\rm ref}$.
We fix the two freedoms by making $\psi-\psi_{\rm ref}$ orthogonal
to these directions in real field coordinates,
\[
\operatorname{Re}\big[(\psi-\psi_{\rm ref})^\dagger F_{\rm ref}\big]=0,
\qquad
\operatorname{Im}\big[\psi_{\rm ref}^\dagger\psi\big]=0.
\]
The second expression uses
$\operatorname{Re}[(\psi-\psi_{\rm ref})^\dagger\ii\psi_{\rm ref}]
=\operatorname{Im}[\psi_{\rm ref}^\dagger\psi]$.
These conditions choose a representative with a fixed time origin and
common phase, without changing the physical orbit. Newton shooting
adjusts all $4L$ real field components, $T$, and $\theta$ to solve
them together with Eq.~\eqref{sm:eq:shooting}.
The converged $\theta$ is zero on the branch studied here, so the full
field repeats after $T$, not just its intensity.

To determine stability, we write the $4L$ real and imaginary field
components as $\bm z$, with equations $\dot{\bm z}=\bm f(\bm z)$.
Along the periodic solution $\bm z_0(t)$, small deviations obey
$\delta\dot{\bm z}=J(t)\delta\bm z$, where
$J_{ij}(t)=\left.\partial f_i/\partial z_j\right|_{\bm z_0(t)}$.
The matrix $B(t)$ maps an initial deviation to its later value,
$\delta\bm z(t)=B(t)\delta\bm z(0)$. It starts from the identity,
since at $t=0$ it maps each deviation to itself, and obeys
\[
\dot B(t)=J(t)B(t),\qquad B(0)=\Id_{4L},\qquad \mathcal M=B(T),
\]
where $\Id_{4L}$ is the identity matrix. The one-period matrix
$\mathcal M$ applies on every circuit because the orbit and $J(t)$
repeat after $T$. Its eigenvalues are the Floquet multipliers.
For $\mathcal M\bm v=\Lambda\bm v$, a deviation along $\bm v$ is
multiplied by $\Lambda$ each circuit, shrinking if $|\Lambda|<1$
and growing if $|\Lambda|>1$.

\begin{figure}[!htbp]
\includegraphics[width=.8\linewidth]{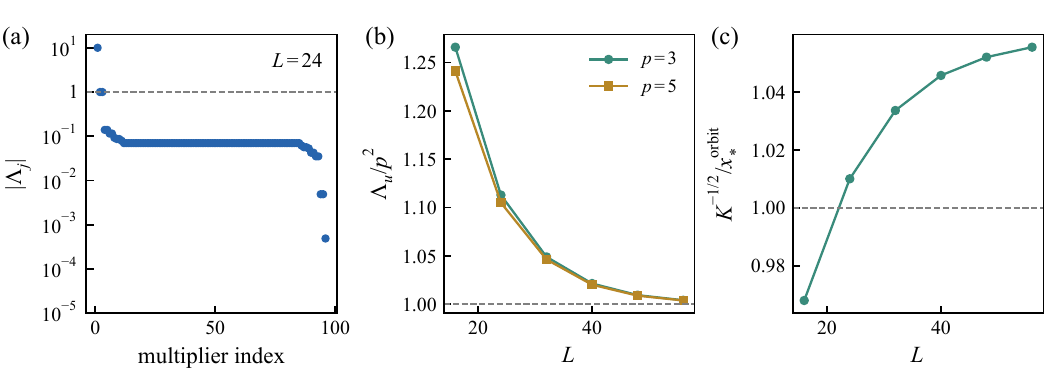}
\caption{Critical circulation and the response prediction. (a) Complete Floquet magnitudes at $L=24$. (b) Unstable multiplier divided by the predicted value $p^2$ for cubic and quintic conversion. (c) Predicted circulating endpoint amplitude divided by the independently computed orbit amplitude.}
\label{sm:fig:orbit}
\end{figure}

The equations are unchanged by shifting the time origin or rotating
all fields by a common phase. For the periodic field $\psi_0(t)$,
these operations produce small deviations proportional to
$\dot\psi_0(t)$ and $\ii\psi_0(t)$, respectively.
Both repeat after $T$ and give unit multipliers. They shift the timing
or common phase of the same motion, so stability is determined by
the remaining modes.
The computed spectrum has one growing mode, with $\Lambda_u>1$,
while all other nonneutral modes decay [Fig.~\ref{sm:fig:orbit}(a)].

Initial fields that approach the orbit, up to time and common phase
shifts, form its stable manifold. The single growing direction makes
this a surface with one fewer dimension than the full initial-condition
space. Varying $A_0$ in $a_1(0)=A_0$, with all other fields initially zero,
traces a line through this space, and $\Ac$ marks its crossing of the
manifold. In full-lattice simulations, seeds just below and above
$\Ac$ approach the orbit before decaying or igniting, respectively.
These trajectories identify the manifold as the local boundary
between the two outcomes.

The peak map $x'=K^4x^9$ in Eq.~\eqref{sm:eq:peak_map} predicts growth
over a complete circuit with two conversions. Differentiating at its
circulating peak $x_*=K^{-1/2}$ gives
\begin{equation}
\Lambda_{\rm map}=\left.\frac{d(K^4x^9)}{dx}\right|_{x=K^{-1/2}}=9.
\end{equation}
A small peak deviation obeys $\delta x'\simeq\Lambda_{\rm map}\delta x$.
Along the full-field unstable mode, the peak deviation grows by a factor
$\Lambda_u$ over the same circuit. The two calculations can therefore
be compared directly, with nine as the return map's approximate
prediction and $\Lambda_u$ obtained independently from the full dynamics.
The coefficient $K$ cancels, leaving a prediction fixed by the
conversion order. For the order-$p$ conversion of
Eq.~\eqref{sm:eq:p_gainlaw}, two successive returns give
$x'=K_p(K_px^p)^p=K_p^{p+1}x^{p^2}$.
Using $K_px_*^{p-1}=1$, the derivative at its nonzero fixed point is
$p^2K_p^{p+1}x_*^{p^2-1}=p^2$.
The full-lattice multipliers approach these values for cubic and
quintic conversion as length increases [Fig.~\ref{sm:fig:orbit}(b)].

The amplitude and multiplier test different aspects of the return map.
The orbit's endpoint peaks remain close to the predicted
$x_*=K^{-1/2}$ [Fig.~\ref{sm:fig:orbit}(c)], whose coefficient $K$
depends on packet shape and distributed conversion. This dependence
cancels from the derivative at the fixed point. The multiplier can
therefore approach nine even when the peak-amplitude ratio remains
different from unity and the circulating packet differs in shape
from the first arriving packet.

\paragraph{Why the departure time grows logarithmically.}

For a seed $A_0=\Ac(1+\varepsilon)$, $\varepsilon$ measures the relative
distance from threshold. After the trajectory approaches the orbit,
the decaying perturbation modes diminish and the growing mode
determines when it leaves.
The single-site family crosses the boundary with a growing-mode
coefficient $d_0=c_1\varepsilon+O(\varepsilon^2)$,
where $c_1\ne0$ measures its sensitivity to the seed amplitude.
After $m$ circuits its magnitude is
\begin{equation}
|d_m|\simeq\Lambda_u^m|d_0|.
\end{equation}
If departure is recorded at a fixed deviation $d_{\rm exit}$,
the condition $|d_m|=d_{\rm exit}$ gives
$m\simeq\ln(d_{\rm exit}/|d_0|)/\ln\Lambda_u$.
Since each circuit takes time $T$, the departure time satisfies
\begin{equation}
\frac{t_{\rm dep}}{T}=c_0+\frac{\ln(1/|\varepsilon|)}{\ln\Lambda_u}+O(1).\label{sm:eq:departure}
\end{equation}
The offset $c_0$ includes the initial approach and the chosen departure
criterion. Recording departure only at discrete arrival peaks gives
the bounded correction.
In the Letter, departure is the first peak of $|a_L|^2$ after $1.5T$
outside the interval from one-half to twice the orbit peak intensity.
Changing this criterion shifts the offset but not the leading
logarithmic slope $1/\ln\Lambda_u$.
The long delay therefore reflects extra circuits near the unstable
orbit, with the slope fixed by its independently computed multiplier.

\section{Sustained motion and its power source}\label{sm:sec:chaos}

\paragraph{Variational dynamics and the chaotic state.}

\begin{figure}[!htbp]
\includegraphics[width=.8\linewidth]{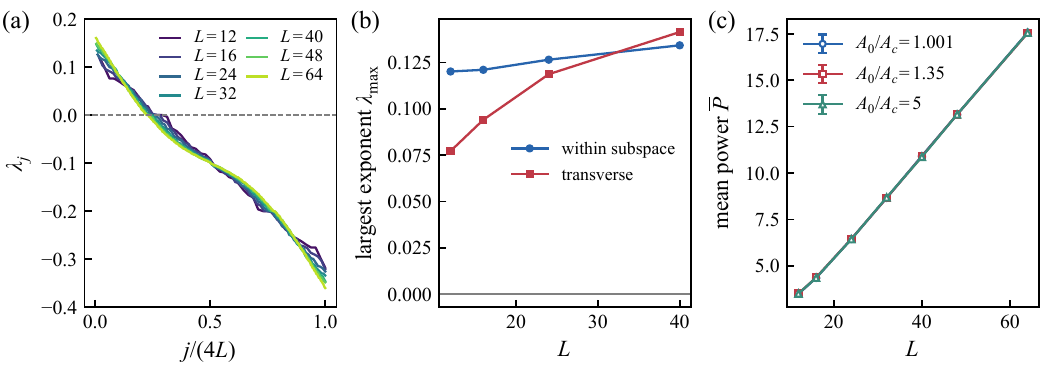}
\caption{Chaotic motion and its power from single-site seeds. (a) Full Lyapunov spectra at $A_0=1.35\Ac$. (b) Largest exponents within and transverse to the exact preparation subspace. (c) Long-time power at three representative amplitudes. Error bars are standard errors of time-block averages.}
\label{sm:fig:chaos}
\end{figure}

The critical orbit describes the transient near threshold. The
long-time state is instead characterized by its Lyapunov spectrum.
For infinitesimal deviations $\delta\bm a$ and $\delta\bm b$ from a
trajectory $(\bm a(t),\bm b(t))$, linearizing Eq.~\eqref{sm:eq:model} gives
\begin{align}
\delta\dot{\bm a}&=M_A\delta\bm a-\ii\chi(\bm I\cdot\delta\bm b+\delta\bm I\cdot\bm b),\\
\delta\dot{\bm b}&=M_B\delta\bm b-\ii\chi(\bm I\cdot\delta\bm a+\delta\bm I\cdot\bm a),\qquad
\delta I_n=2\operatorname{Re}(a_n^*\delta a_n+b_n^*\delta b_n).\label{sm:eq:tangent}
\end{align}
Here $\delta I_n$ is the intensity variation. Because it contains
complex conjugates, we solve these linear equations in the $4L$ real
and imaginary field components. A basis of perturbations is evolved
alongside the trajectory and periodically orthonormalized by QR
decomposition. The accumulated logarithms of the stretching factors,
divided by elapsed time, give the Lyapunov exponents $\lambda_j$.
After a transient of $5000$ time units and a further $1000$ for the
perturbation basis to align, we accumulate the spectrum over $6000$
time units, with QR steps every $0.2$ and integration step $0.02$.
This auxiliary calculation leaves the underlying single-site trajectory
unchanged. Positive exponents show that nearby trajectories separate
exponentially and confirm chaos at $A_0=1.35\Ac$
[Fig.~\ref{sm:fig:chaos}(a)].

Hopping and conversion preserve phase-space volume, while loss
contracts each real coordinate at rate $\Gamma$.
Their total contraction fixes $\sum_{j=1}^{4L}\lambda_j=-4\Gamma L$,
providing a check on the spectrum.

At the reference parameters, positive exponents occur for perturbations
both within and outside the real--imaginary preparation subspace of
Sec.~\ref{sm:sec:methods} [Fig.~\ref{sm:fig:chaos}(b)]. The motion is chaotic
within this subspace and also has growing transverse perturbations.
Across six seed amplitudes spanning $1.001\le A_0/\Ac\le5$ and seven
lengths from $L=12$ to $64$, the mean power varies little with seed
amplitude and grows approximately linearly with length
[Fig.~\ref{sm:fig:chaos}(c)]. We average over
$5\times10^3\le t\le4.5\times10^4$, using blocks of duration $10^3$
to estimate statistical error. Twenty longer runs at
$L=8,10,12,16,24$ remain active to $t=2\times10^5$, with positive
largest Lyapunov exponents in the final measurement windows.

\paragraph{Power balance and exact decay controls.}

The source of this sustained motion can be identified from its power
balance. Conversion at any real local rate $q_n$ makes equal and
opposite contributions to the two intensities,
\begin{equation}
\left.\frac{d|a_n|^2}{dt}\right|_{\rm conv}=2q_n\operatorname{Im}(a_n^*b_n)
=-\left.\frac{d|b_n|^2}{dt}\right|_{\rm conv}.
\end{equation}
Conversion therefore conserves total power $P=\sum_n I_n$.
Adding hopping and loss gives
\begin{equation}
\dot P=\mathcal J_{\rm NH}-2\Gamma P,\qquad
\mathcal J_{\rm NH}=4\gamma\sum_{n=1}^{L-1}
\operatorname{Im}(b_n^*b_{n+1}-a_n^*a_{n+1}).\label{sm:eq:power}
\end{equation}
The term $\mathcal J_{\rm NH}$ is the power supplied by active
nonreciprocal hopping. In statistically stationary motion, this supply
balances loss,
$\langle\mathcal J_{\rm NH}\rangle=2\Gamma\langle P\rangle$.
Hopping therefore powers the motion after the seed has initiated it,
while conversion returns the field to an amplifying channel.

Sufficiently strong loss prevents sustained motion. Since conversion conserves
total power, differentiating $P=\bm a^\dagger\bm a+\bm b^\dagger\bm b$
leaves only the hopping and loss matrices
$M_s=-\ii H_s-\Gamma\Id_L$, with $s=A,B$,
\[
\dot P=\bm a^\dagger(M_A+M_A^\dagger)\bm a
       +\bm b^\dagger(M_B+M_B^\dagger)\bm b.
\]
Thus instantaneous power growth is governed by the Hermitian matrices
$Q_s=(M_s+M_s^\dagger)/2$. For chain $A$, $Q_A$ has diagonal entries
$-\Gamma$ and nearest-neighbor entries $\ii\gamma$ and $-\ii\gamma$.
A site-dependent phase change makes both hopping entries real and
equal to $\gamma$, without changing the eigenvalues. Its spectrum is
therefore that of a reciprocal open chain shifted by $-\Gamma$.
Chain $B$ has the same spectrum, with largest eigenvalue
$\mu_{\max}=2\gamma\cos[\pi/(L+1)]-\Gamma$.
For any field $\bm x$, $\bm x^\dagger Q_s\bm x\le
\mu_{\max}\|\bm x\|^2$, so
\begin{equation}
\dot P\le2\left(2\gamma\cos\frac{\pi}{L+1}-\Gamma\right)P.
\label{sm:eq:decay_bound}
\end{equation}
The coefficient on the right is strictly negative for every finite
chain when $\Gamma\ge2\gamma$, so all fields decay regardless of
their initial amplitude. For reciprocal hopping, $\gamma=0$, the hopping
contribution vanishes and $\dot P=-2\Gamma P$ exactly.
The corresponding bulk field-growth rate $2\gamma-\Gamma$ is the
largest imaginary part of the periodic-boundary dispersions
$E_{A,B}(k)=2\cos k-\ii\Gamma\mp2\ii\gamma\sin k$.
Positive bulk growth coincides with $r_\infty>0$, allowing directional
amplification despite open-boundary spectral
stability~\cite{Wanjura2020,Midya2022}.

With weaker loss, amplification is possible, but gain alone need not
sustain motion. To test this, we orient both chains the same
way, setting $H_B=H_A$.
They share the positive weight matrix $W=S^{-2}$, with
$W_{nn}=e^{-2\kappa(n-1)}$ and $WH_A=H_A^\dagger W$.
Hopping therefore conserves the weighted power, as does conversion
between equally weighted sites. Hence
\begin{equation}
P_W=\bm a^\dagger W\bm a+\bm b^\dagger W\bm b,
\qquad \dot P_W=-2\Gamma P_W.\label{sm:eq:same_skin}
\end{equation}
Positive weights ensure that every finite initial field decays.
Reflection of one uncoupled chain is unitary, so this control retains
the same resting-state spectrum and maximum linear transient gain in
total power. Yet conversion no longer feeds an amplifying return path.

\paragraph{A linear return path with onsite Kerr nonlinearity.}

The return path can also be made linear by replacing $q_n=\chi I_n$
with a constant $\delta$. Adding onsite Kerr terms then gives a
nonlinear lattice with a different feedback mechanism,
\begin{equation}
\dot{\bm a}=M_A\bm a-\ii\delta\bm b-\ii U|\bm a|^2\cdot\bm a,\qquad
\dot{\bm b}=M_B\bm b-\ii\delta\bm a-\ii U|\bm b|^2\cdot\bm b.
\label{sm:eq:kerr_control}
\end{equation}
Here $\chi=0$, $\delta$ is real, and $U$ is the real Kerr coefficient.
Kerr changes local phases without directly changing intensities,
while $\delta$ transfers intensity between chains.

To establish decay at weak coupling, write $\bm a=S_c\bm y$ and
$\bm b=S_c^{-1}\bm z$, with
$(S_c)_{nn}=e^{\kappa[n-(L+1)/2]}$. This centered gauge makes both
intrachain hopping matrices Hermitian.
The weighted power $N_W=\|\bm y\|^2+\|\bm z\|^2$ is conserved by
hopping and Kerr phase shifts. Linear conversion does not cancel
under the opposite weights, giving
\begin{align}
\dot N_W
&=-2\Gamma N_W+
2\delta\,\operatorname{Im}\!\left[\bm y^\dagger(S_c^{-2}-S_c^2)\bm z\right]\notag\\
&\le-2\{\Gamma-|\delta|\sinh[\kappa(L-1)]\}N_W.
\label{sm:eq:kerr_bound}
\end{align}
The inequality follows from
$\|S_c^{-2}-S_c^2\|=2\sinh[\kappa(L-1)]$ and
$2\|\bm y\|\|\bm z\|\le N_W$.
For the diagonal matrix $S_c^{-2}-S_c^2$, this norm is its largest
absolute diagonal entry.
Loss dominates for
$|\delta|<\Gamma/\sinh[\kappa(L-1)]$, forcing every initial field
to decay for any real $U$.

For positive Kerr coefficients, the rescaling
$(\widetilde{\bm a},\widetilde{\bm b})
=\sqrt{U}(\bm a,\bm b)$ reduces the equations to $U=1$,
which we use below.
We also test linearly stable cases beyond the sufficient decay bound at
$L=24$, with $\delta=0.005$--$0.0099$ and the reentrant
stable region $\delta=0.437$--$0.48$.
All tested single-site seeds between $A_0=10^{-3}$ and $3$ decay.
The sustained states found in this control appear after the resting
state becomes linearly unstable. They are coherent at
$\delta=0.011,0.015$ and chaotic at $\delta=0.02,0.1$, and are
reached from seeds down to $A_0=10^{-12}$.
Smaller seeds delay onset by $\sigma^{-1}\ln(1/A_0)$, where $\sigma>0$ is the
linear growth rate. In both chaotic cases, the long-time mean power
shows no systematic growth in time and varies little with seed
amplitude at fixed parameters. The sustained motion found in this
control thus follows linear instability, rather than bypassing it.

\paragraph{Conversion with additional onsite phase shifts.}

To test Kerr phase shifts without replacing the nonlinear return path,
we retain $q_n=\chi I_n$ and add $-\ii U|a_n|^2a_n$ and
$-\ii U|b_n|^2b_n$, with $\delta=0$.
At the reference parameters and $L=24$, $U=0.5,2$ shift the threshold
by less than $0.05\%$ and pass the step and duration checks of
Sec.~\ref{sm:sec:methods}. Both give sustained chaos at $A_0=1.35\Ac$,
with positive largest Lyapunov exponents.
The leading return response is unchanged because Kerr first corrects
the field in chain $A$ at order $A_0^3$. In the conversion source
$|a_n|^2a_n$, this correction is multiplied by two fields of order
$A_0$, giving a change only at order $A_0^5$. The source that generates
the leading $A_0^3$ response in chain $B$ is therefore unaffected.

\section{Noise-induced ignition}\label{sm:sec:noise}

\paragraph{Noise and ignition protocol.}

Noise can provide the initial disturbance without a prepared seed.
Starting from the resting state, we add independent random fields
at every site in both chains. After each deterministic RK4 step
of duration $\Delta t=0.04$, every complex amplitude receives
the increment
\[
\Delta\psi_j^{\rm noise}
=\sqrt{D\Delta t}\,(\xi_{j,R}+\ii\xi_{j,I}),
\]
where $D$ is the noise strength and the $\xi$ are independent
Gaussian variables with zero mean and unit variance, drawn anew
for every field and time step. The real and imaginary parts
therefore have equal noise strength.

In continuous time, this prescription is written as
\begin{equation}
\begin{aligned}
d\psi_j&=f_j(\psi)\,dt+\sqrt{2D}\,dW_j,\\
\langle dW_jdW_k^*\rangle&=\delta_{jk}\,dt,
\qquad \langle dW_jdW_k\rangle=0.
\end{aligned}
\label{sm:eq:noise}
\end{equation}
Here $\psi=(\bm a,\bm b)$ contains the $2L$ field amplitudes and
$f_j$ is the deterministic rate of change given by Eq.~\eqref{sm:eq:model}.
The complex Wiener increment $dW_j$ has zero mean, and brackets denote
averaging over noise realizations. Its first correlation specifies
the variance and independence between fields. The second gives equal,
uncorrelated real and imaginary parts. Increments over disjoint time
intervals are independent as well.

We monitor ignition during an observation window $t_{\rm obs}$,
with noise applied continuously until ignition or the end of
the window. Ignition is recorded when the total power first
exceeds $P=1$, checking once per unit time. In the tested range,
this level lies above the critical-orbit power and below the
sustained-state power. We evaluate ignition probabilities at
$t_{\rm obs}=10^3,10^4,10^5$, using 128 independent realizations
at each sampled length and noise strength.
Runs that have not ignited are followed to the corresponding
$t_{\rm obs}$, including the longest window at $10^5$.
We use binomial Wilson confidence intervals and interpolate between
adjacent lengths to obtain $L_{1/2}$, where half the realizations ignite.

We check time-step convergence by halving $\Delta t$ from
$0.04$ to $0.02$ using the same 128 noise realizations.
The ignition probabilities and half-ignition lengths agree
within sampling uncertainty at all three observation times.

A separate test checks whether crossing the power threshold marks
self-sustained chaos. We continue 48 crossing states at
$L=16,24,32,48$ for 6000 further time units, once with noise left
on and once with it switched off at the crossing. All retain positive
largest Lyapunov exponents over the final 4000 units, and halving the
step preserves these outcomes. One noise-free trajectory per length
remains chaotic to $t=20000$. The motion therefore survives removal
of the noise that triggered it.

\paragraph{Amplification of noise.}

Before ignition, the noise-induced field can be estimated from linear
propagation, neglecting conversion. Consider noise injected at site $j$
of chain $A$ over a short interval $\mathrm{d}s$ around time $s$.
This increment has variance $2D\,\mathrm{d}s$, and its field is multiplied
by $[U_A(t-s)]_{Lj}$ on reaching the right end at time $t$.
Here $U_A$ is the propagator defined in Sec.~\ref{sm:sec:return}.
Its contribution to the endpoint variance is consequently
$2D|[U_A(t-s)]_{Lj}|^2\,\mathrm{d}s$.
Independent contributions from different sites and times add, giving
\begin{equation}
\langle |a_L(t)|^2\rangle_{\rm lin}
=2D\sum_{j=1}^{L}\int_0^t
|[U_A(t-s)]_{Lj}|^2\,\mathrm{d}s.
\label{sm:eq:noise_sources}
\end{equation}
The subscript denotes linear evolution, for which the mean field is
zero and the averaged intensity equals the variance. Since each
increment eventually decays, the variance approaches a finite
long-time value.

The variance contains the full time-integrated response from every
injection site, whereas $G$ measures the peak response from site $1$.
Writing $\tau=t-s$ for the elapsed propagation time separates the
peak gain from the remaining integral,
\[
\begin{aligned}
\langle |a_L|^2\rangle_{{\rm lin},\infty}
&=2DG^2\,\tau_{\rm eff}(L),\\
\tau_{\rm eff}(L)
&\equiv\sum_{j=1}^{L}\int_0^\infty
\left|\frac{[U_A(\tau)]_{Lj}}{G}\right|^2\,\mathrm{d}\tau.
\end{aligned}
\]
The subscript $\infty$ denotes the long-time limit, and
$\tau_{\rm eff}$ collects the response duration and the relative
contributions of the injection sites.

For long chains with $r_\infty>0$, noise near the left end
dominates the right-end fluctuations because it traverses the
longest amplified path.
The propagation estimate in Sec.~\ref{sm:sec:return} gives
$G=e^{r_\infty L+o(L)}$.
Pulse widths and the sum over injection sites change more slowly
with length and contribute no additional exponential factor,
so $\tau_{\rm eff}=e^{o(L)}$.
Consequently,
\[
\langle |a_L|^2\rangle_{{\rm lin},\infty}
=DG^2e^{o(L)}
=D e^{2r_\infty L+o(L)}.
\]
The squared gain therefore sets the leading length dependence.
Mirror symmetry gives the same result for $b_1$, where fluctuations
generated near the right end of chain $B$ undergo the longest
amplified passage.

\paragraph{The logarithmic length scale.}

These amplified fluctuations must be compared with the circulating
peak $x_*=\pred G$. The required intensity scales as
$\pred^2G^2$, while the noise variance scales as $DG^2$.
Their ratio therefore scales as $\pred^2/D$, with a larger ratio
requiring a rarer fluctuation to trigger motion.

We estimate the ignition probability at fixed $t_{\rm obs}$ as a
function of this ratio, treating changes in the spatial structure,
duration, and recurrence of fluctuations as subleading corrections.
Within this scaling approximation, half-ignition corresponds to a
fixed $\pred^2/D$. Using $\pred^2\propto\chi^{-1}e^{-3r_\infty L}$
from Eq.~\eqref{sm:eq:asymptote} then gives
\begin{equation}
L_{1/2}(D;t_{\rm obs})\simeq\frac{1}{3r_\infty}
\ln\frac{1}{\chi D}+C(t_{\rm obs}).
\label{sm:eq:half_length}
\end{equation}
The offset $C(t_{\rm obs})$ depends on the observation time and noise
protocol. Packet spreading and fluctuation duration introduce subleading
finite-length corrections to this leading logarithmic law.
As in the deterministic problem, the rescaling
$\widetilde\psi=\sqrt\chi\psi$ sets the conversion strength to one.
It also changes the noise strength to $\chi D$ and the detection
level to $\widetilde P=\chi$.

In the noise simulations, the fitted slopes of $L_{1/2}$ versus
$\log_{10}(1/D)$ are
$5.18$, $5.20$, and $5.14$ sites per decade for
$t_{\rm obs}=10^3,10^4,10^5$, respectively.
Their near equality across the observation windows and agreement
with the leading prediction $\ln10/(3r_\infty)\simeq5.06$
support the predicted logarithmic dependence.

\section{Periodic motion and dimerized chains}\label{sm:sec:ssh}

\paragraph{A regular sustained state in the main model.}

As loss approaches the amplification boundary $\Gamma=2\gamma$ from below, the sustained motion in the uniform chains can become periodic. At $(\gamma,\Gamma)=(0.2,0.34)$ and $L=40$, the threshold is $\Ac\simeq0.584$, and a single-site seed $A_0=1.2\Ac$ produces oscillating local intensities.

\begin{figure}[t]
\centering\includegraphics[width=.8\linewidth]{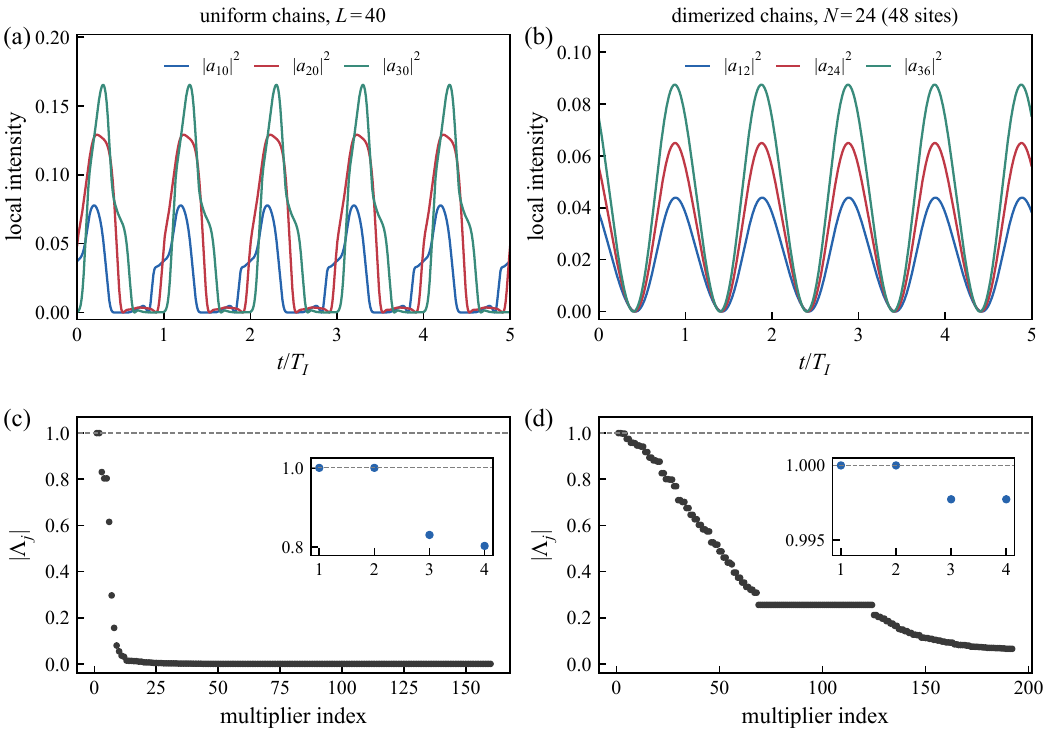}
\caption{Stable periodic motion from a single-site seed $A_0=1.2A_{\mathrm c}$. (a,b) Local intensities over five intensity periods $T_I$ after transients, with the time origin reset. Subscripts label physical sites, with $2N=48$ sites per chain in (b). Here $T_I=T$ in (a) and $T_I=T/2$ in (b), where $T$ is the field period. (c,d) Floquet multiplier magnitudes over $T$. All lie below unity except the two symmetry-neutral multipliers, confirming stability of the periodic states. Insets enlarge the leading multipliers. The uniform example has $\Gamma=0.34$, $L=40$, and the dimerized example has $\Gamma=0.24$, $N=24$, $(v,w)=(1.3,0.7)$.}
\label{sm:fig:regular}
\end{figure}

The field and local intensities repeat after $T\simeq40.77723$.
Varying the single-site seed amplitude over five values between
$1.01A_{\mathrm c}$ and $2A_{\mathrm c}$ yields periodic motion
with the same period and Floquet spectrum [Fig.~\ref{sm:fig:regular}(a,c)].
The one-period matrix $\mathcal M$, defined in Sec.~\ref{sm:sec:orbit},
has two neutral multipliers associated with time and global phase shifts.
All others have modulus at most $0.83121$, establishing stability also
outside the preparation subspace. Halving the time step leaves these
values unchanged at the displayed precision.

\paragraph{A nearest-neighbor dimerized extension.}

\begin{figure}[b]
\centering\includegraphics[width=.8\linewidth]{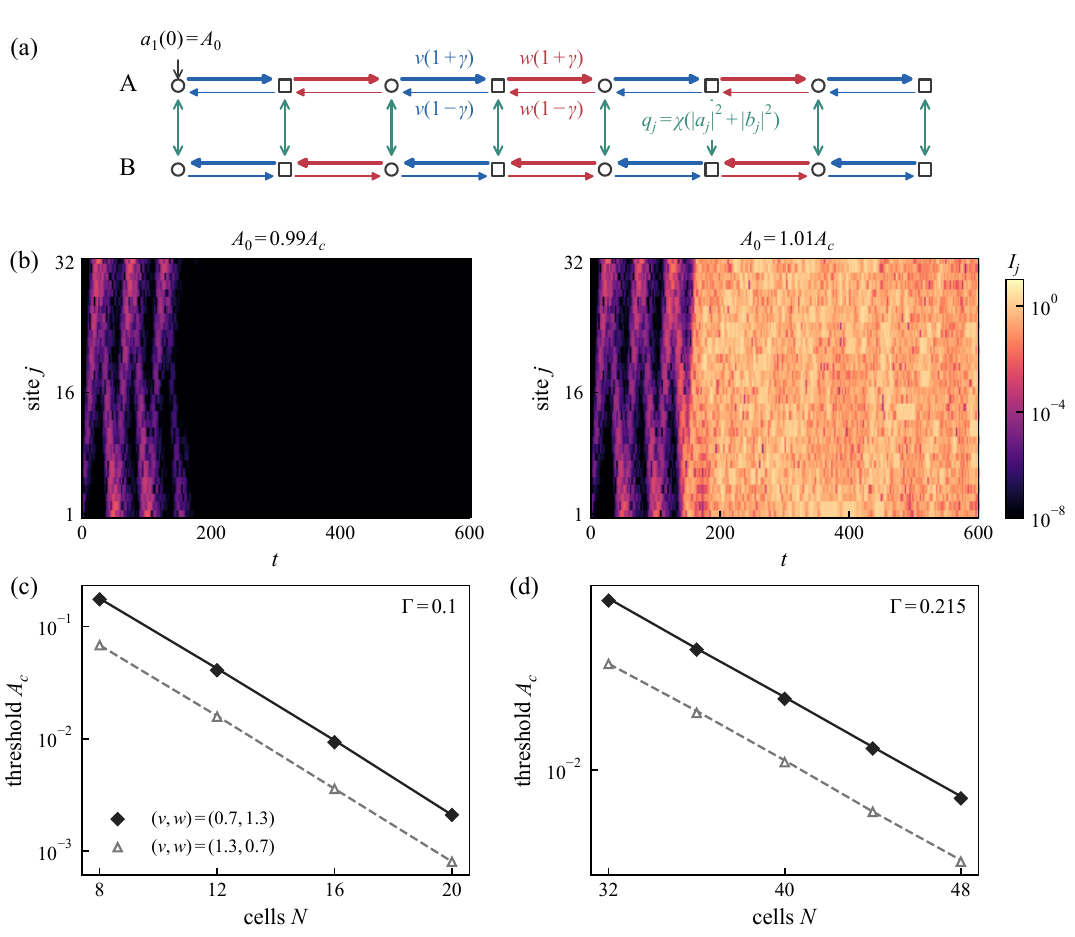}
\caption{Triggering in dimerized chains. (a) Alternating bonds $v,w$ have stronger hopping $(1+\gamma)t_j$ along the thick arrows and weaker hopping $(1-\gamma)t_j$ in reverse. Green rungs denote local intensity-dependent conversion. (b) Site intensity $I_j(t)$ below and above threshold at $N=16$, $(v,w)=(1.3,0.7)$, and $\Gamma=0.1$, with a common logarithmic color scale. (c,d) Single-site thresholds (symbols) and endpoint predictions (lines) for two losses. Filled diamonds and open triangles distinguish the two dimerizations. Here $\gamma=0.2$, $\chi=1$, and each chain has $N$ cells or $2N$ sites.}
\label{sm:fig:dimer}
\end{figure}

Replacing uniform hopping by alternating bonds lets us test the return condition in a two-band lattice. Each chain contains $N$ two-site cells, or $\ell=2N$ physical sites. The fields $a_j,b_j$ are indexed by site, so cell $m$ contains sites $2m-1$ and $2m$. The alternating bonds are
\begin{equation}
t_{2m-1}=v,\qquad t_{2m}=w,\qquad
(H_A)_{j+1,j}=(1+\gamma)t_j,\quad
(H_A)_{j,j+1}=(1-\gamma)t_j,\quad H_B=H_A^{\mathsf T}.
\label{sm:eq:dimer_h}
\end{equation}
The local conversion remains $q_j=\chi(|a_j|^2+|b_j|^2)$, and only $a_1(0)=A_0$ is initially nonzero. Figure~\ref{sm:fig:dimer}(a) shows this lattice, and Fig.~\ref{sm:fig:dimer}(b) illustrates decay and sustained motion on opposite sides of its triggering threshold.

With $S_{jj}=e^{\kappa(j-1)}$,
\begin{equation}
S^{-1}H_AS=hH_{\rm SSH}(v,w),\qquad h=\sqrt{1-\gamma^2},
\label{sm:eq:dimer_gauge}
\end{equation}
where $H_{\rm SSH}$ is the real symmetric Su--Schrieffer--Heeger hopping matrix~\cite{SuSchriefferHeeger1979}. Every resting-state mode therefore decays at exactly $\Gamma$, for every $N$. The two dimerizations are topological ($v<w$) and trivial ($v>w$) in this reciprocal representative. Both support triggering. The uniform case $v=w=1$ recovers the main model with $L=2N$.

\paragraph{Threshold prediction and periodic output.}

Using the dimerized hopping matrices in Eqs.~\eqref{sm:eq:linear_aux} and \eqref{sm:eq:cubic_aux} gives the arrival at $a_\ell$ and return at $b_1$. Their peak coefficients $G,Y$ enter the same prediction $\pred=\sqrt{G/Y}$, without fitted coefficients. We extract the first arrival and return over $[0.35,1.8]\tau_\ell$ and $[1,3.5]\tau_\ell$, respectively, with reference time $\tau_\ell=(\ell-1)/(2h\min(v,w))$.

For $(v,w)=(0.7,1.3)$ and $(1.3,0.7)$, the predictions agree with the numerical thresholds to within $4\%$ at $\Gamma=0.1$ and $2.4\%$ at $\Gamma=0.215$ [Fig.~\ref{sm:fig:dimer}(c,d)]. The latter set includes $N=32,36,40,44,48$. The threshold brackets and the two trajectories in Fig.~\ref{sm:fig:dimer}(b) retain their outcomes under the duration and step-size checks of Sec.~\ref{sm:sec:methods}.

The dimerized chains also support regular motion near their amplification boundary. For both dimerizations, the periodic-boundary spectrum of Eq.~\eqref{sm:eq:dimer_h} has maximum imaginary part approximately $0.286$ before loss is added. Increasing loss toward this value can produce stable periodic motion, as in the uniform chains.

\begin{samepage}
For $(v,w)=(1.3,0.7)$, $\Gamma=0.24$, and $N=24$, the numerical threshold is $\Ac\simeq0.373$. A single-site seed $A_0=1.2\Ac$ reaches
\begin{equation}
\psi(t+T)=\psi(t),\qquad T\simeq5.68506.
\label{sm:eq:dimer_period}
\end{equation}
\end{samepage}
The field changes sign after $T/2$, giving an intensity period $T_I=T/2$, whereas $T_I=T$ for the uniform example. The one-period matrix $\mathcal M=D\Phi_T$ has two neutral multipliers and all other moduli at most $0.99774$. Varying the single-site seed amplitude over five values between $1.01A_{\mathrm c}$ and $2A_{\mathrm c}$ leads to the same periodic motion. The one-period matrices for both examples were also checked against finite differences and a halved integration step. Both periodic states therefore attract nearby trajectories, including those outside the preparation subspace [Fig.~\ref{sm:fig:regular}(c,d)]. This distinguishes them from the critical orbit at threshold, whose growing mode drives nearby trajectories away.

\section{Residual linear coupling}\label{sm:sec:delta}

\paragraph{The stable length window.}

Adding a real constant coupling gives $q_n=\delta+\chi I_n$.
Unlike the intensity-dependent conversion, $\delta$ remains in
the linearization about the resting state. The linear equation is
$\dot\psi=J(\delta)\psi$, with $\psi=(\bm a,\bm b)$ and
\begin{equation}
J(\delta)=-\Gamma\Id_{2L}-\ii
\begin{pmatrix}H_A&\delta\Id_L\\\delta\Id_L&H_B\end{pmatrix}.
\end{equation}
This linear return path makes the spectrum length dependent,
as in the critical NHSE~\cite{LiCritical2020,YiYang2025}.

Stability is first lost when a linear mode stops decaying.
Write this mode as $(\bm a(t),\bm b(t))=
(\bm a_\omega,\bm b_\omega)e^{-\ii\omega t}$, where $\omega$
is real and $\bm a_\omega,\bm b_\omega$ are its spatial field
profiles on the two chains. Substitution gives
\[
(z\Id_L-H_A)\bm a_\omega=\delta\bm b_\omega,\qquad
(z\Id_L-H_B)\bm b_\omega=\delta\bm a_\omega,\qquad
z=\omega+\ii\Gamma.
\]
Each chain is driven by the field in the other. With response matrix
$\mathcal G_s(z)=(z\Id_L-H_s)^{-1}$, the two relations become
$\bm a_\omega=\delta \mathcal G_A\bm b_\omega$ and
$\bm b_\omega=\delta \mathcal G_B\bm a_\omega$.
Substituting the second relation into the first gives
$\bm a_\omega=\delta^2\mathcal G_A\mathcal G_B\bm a_\omega$.
A nonzero field can satisfy this equation only if
$\Id_L-\delta^2\mathcal G_A\mathcal G_B$ is singular, hence
\begin{equation}
\det[\Id_L-\delta^2\mathcal G_A(z)\mathcal G_B(z)]=0.\label{sm:eq:linear_loop}
\end{equation}
The field must therefore return with both its amplitude and phase
reproduced. The first instability occurs at the smallest $|\delta|>0$
satisfying this condition for real $\omega$, denoted by $|\delta_c(L)|$.
Direct diagonalization confirms the crossing at $\omega=0$ for the
reference even-length chains.

For $r_\infty>0$, each directed leg has a zero-frequency gain
scaling as $e^{r_\infty L}$ (Sec.~\ref{sm:sec:return}).
Two propagation legs and two conversions therefore give a loop
gain scaling as $\delta^2e^{2r_\infty L}$. Setting it to unity gives
\begin{equation}
|\delta_c(L)|\sim C_\delta e^{-r_\infty L},\qquad
L_\delta\simeq r_\infty^{-1}\ln\frac{C_\delta}{|\delta|}.\label{sm:eq:linear_length}
\end{equation}
Here $C_\delta$ is a prefactor set by the channel parameters.
At fixed residual coupling, $|\delta_c(L_\delta)|=|\delta|$ defines
the length $L_\delta$ at which the resting state first loses stability.

\paragraph{The return below the linear crossing.}

Below $|\delta_c(L)|$, the resting state remains linearly stable.
For weak residual coupling, we approximate the field in chain $A$
during its first passage by
$\bm a(t)\simeq A_0\bm a^{(1)}(t)$,
where $\bm a^{(1)}$ is the uncoupled chain's linear response
to a unit excitation at $A_1$, defined in Sec.~\ref{sm:sec:return}.
The constant coupling supplies chain $B$ with the source
$-\ii A_0\delta\bm a^{(1)}(t)$. Factoring out the seed amplitude and
coupling strength, we write the generated field as
$A_0\delta\bm b_{\rm lin}$, where the response $\bm b_{\rm lin}$ obeys
\[
\dot{\bm b}_{\rm lin}=M_B\bm b_{\rm lin}-\ii\bm a^{(1)},\qquad
\bm b_{\rm lin}(0)=0,
\]
where $M_B$ describes propagation and loss in chain $B$.

Cubic conversion generates the additional field $A_0^3\bm b^{(3)}$
from Sec.~\ref{sm:sec:return}. At this order, both fields propagate
linearly in chain $B$ and add coherently,
$\bm b\simeq A_0\delta\bm b_{\rm lin}+A_0^3\bm b^{(3)}$.
Equating the peak magnitude of this sum at $B_1$ to the arrival
peak $A_0G$ at $A_L$, and dividing by $A_0>0$, gives
\begin{equation}
\max_{t\in W_r}|\delta b_{{\rm lin},1}(t)+A_0^2b_1^{(3)}(t)|=G.\label{sm:eq:coherent}
\end{equation}
Here $W_r$ selects the first return. Solving for $A_0$ predicts the
shifted threshold and recovers $\pred=\sqrt{G/Y}$ at $\delta=0$.

\begin{figure}[!htbp]
\centering\includegraphics[width=.8\linewidth]{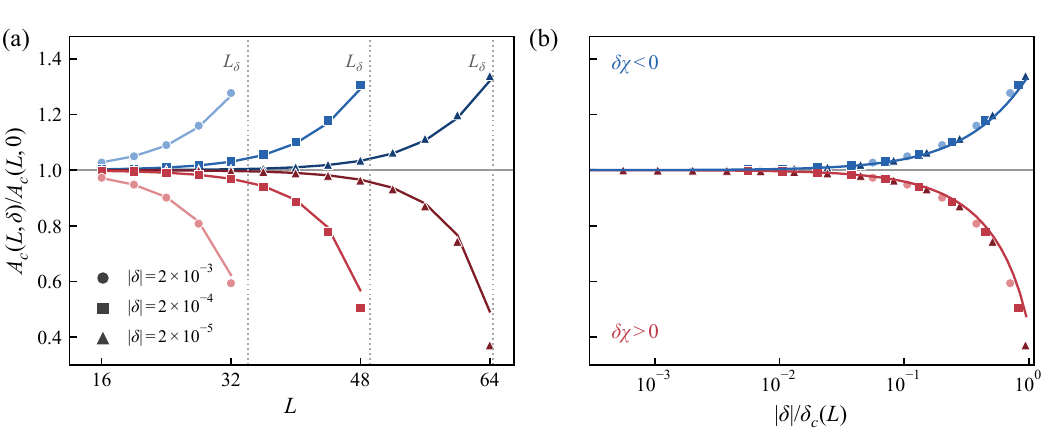}
\caption{Triggering below the residual-coupling instability. (a) Threshold normalized to its value at $\delta=0$. Circles, squares, and triangles denote $|\delta|=2\times10^{-3},2\times10^{-4},2\times10^{-5}$, respectively. Red and blue distinguish positive and negative coupling. Dotted lines indicate the associated critical lengths. (b) The same data versus $|\delta|/\delta_c(L)$. Solid curves use the coherent first-return approximation, Eq.~\eqref{sm:eq:coherent}.}
\label{sm:fig:delta}
\end{figure}

Figure~\ref{sm:fig:delta} compares this prediction with full dynamics.
For small $|\delta|/\delta_c(L)$, the threshold changes approximately
linearly with signed $\delta/\delta_c(L)$, decreasing for positive
coupling and increasing for negative coupling as the two return
fields reinforce or oppose each other. Near the linear crossing,
repeated linear circulation changes the wave packet, and the
numerical thresholds depart from the first-return prediction.

Beyond the first crossing, the resting state is linearly unstable,
so the threshold no longer separates decay from sustained motion.
At the reference parameters, we sampled $L=24,32,40$ and couplings
in the range $1.02\le|\delta|/\delta_c(L)\le2$.
Positive coupling leads to chaos from very small single-site seeds
in these tests. For negative coupling, small seeds reach low-power
nonzero motion and larger seeds can reach chaos.

\end{document}